\documentclass[12pt]{iopart}

\usepackage{graphics}
\usepackage{siunitx}
\usepackage[position=top]{subfig}
\usepackage{multirow}
\usepackage[utf8]{inputenc}
\usepackage{booktabs}
\usepackage{float}
\usepackage{array}
\usepackage{iopams}
\usepackage{pgfplots}
\pgfplotsset{compat=newest}
\usepackage[title]{appendix}
\usepackage{doi}

\newcommand{\be}{\begin{eqnarray}}
\newcommand{\ee}{\end{eqnarray}}
\newcommand{\ket}[1]{\ensuremath{\left| {#1} \right>}}
\newcommand{\bra}[1]{\ensuremath{\left< {#1} \right|}}
\newcommand{\braket}[2]{\ensuremath{\left< \left. {#1} \right| {#2} \right>}}

\newcommand{\nbar}{\bar{n}}
\newcommand{\tf}{t_{\rm f}}
\newcommand{\bmax}{\beta_{\rm max}}
\newcommand{\caf}{\ensuremath{^{40}{\rm Ca}^{+}\, }}
\newcommand{\eqdef}{=\mathrel{\mathop:}}

\DeclareFontFamily{U}{wncy}{}
\DeclareFontShape{U}{wncy}{m}{n}{<->wncyr10}{}
\DeclareSymbolFont{mcy}{U}{wncy}{m}{n}

\begin{document}
\title{Robust dynamical exchange cooling with trapped ions}

\author{T.~Sägesser, R.~Matt, R.~Oswald and J.~P.~Home  }

\address{Institute for Quantum Electronics, ETH Zürich, Otto-Stern-Weg 1, 8093 Zürich, Switzerland}
\ead{tobiass@phys.ethz.ch}


\begin{abstract}
We investigate theoretically the possibility for robust and fast cooling of a trapped atomic ion by transient interaction with a pre-cooled ion. The transient coupling is achieved through dynamical control of the ions' equilibrium positions. To achieve short cooling times we make use of shortcuts to adiabaticity by applying invariant-based engineering. We design these to take account of imperfections such as stray fields, and trap frequency offsets. For settings appropriate to a currently operational trap in our laboratory, we find that robust performance could be achieved down to 6.3 motional cycles, comprising \SI{14.2}{\micro\second} for ions with a \SI{0.44}{\mega\hertz} trap frequency. This is considerably faster than can be achieved using laser cooling in the weak coupling regime, which makes this an attractive scheme in the context of quantum computing.
\end{abstract}

\maketitle

\section{Introduction}
\label{sec:introduction}
One of the major challenges in quantum computing is to realise fast operations, since these affect both the clock speed as well as the ability to preserve coherence in the presence of decoherence mechanisms. For trapped-ion approaches, direct operations on the qubits include single and multi-qubit gates and state detection. However in order to implement these processes in the flexible manner and with the high reliability required for quantum error-correction, transport of ions and re-cooling are expected to play an important role \cite{02Kielpinski, 09Home, Negnevitsky2018}. Thus to increase the speed of a trapped-ion quantum information processor all of these processes must be improved. While recent work has demonstrated impressive progress in the speed of one and two-qubit gates \cite{Schaefer2018}, detection \cite{Cahall19} and transport \cite{11Bowler, 11FSKgroup}, in many recent demonstrations in multi-zone chips the primary speed limitation was due to laser-recooling of ions, either after imperfect transport or following detection events, which heat the ions via photon recoil \cite{Negnevitsky2018, 10Hanneke}. Laser cooling close to the ground state is performed by resolved-sideband cooling, while methods such as Sisyphus cooling \cite{Dalibard85} and electromagnetically-induced transparency cooling \cite{00Roos} offer higher rates. These methods are limited in rate by working in the weak coupling regime and by fundamental features of the atom, including finite decay rates for spontaneous emission of around $10^8$~s$^{-1}$ and the imperfect transfer of momentum between the atom and the light field, leading to cooling cycles of several hundred trap periods in current experiments \cite{Negnevitsky}. While it may be possible to perform laser cooling on timescales faster than the ion oscillation period \cite{Machnes2010}, the recoil rate presents a hard limit.

The premise for our work is that exchange of energy of ions via the Coulomb interaction can be used to extract excess energy from a hot ion by bringing it into resonance with a pre-cooled ion in a nearby potential well. Two ions held close to each other in an external potential experience a mutual repulsive force. This modifies the ion equilibrium positions relative to the minima of the external potential, but also couples the vibrations of the two different ions. For two ions of mass $m_1$ and $m_2$ separated by a distance $d$, which are held in harmonic potential wells in which they oscillate with frequencies $\omega_1$ and $\omega_2$ respectively, energy exchange between the oscillations of each ion occurs at a frequency
\be
\label{eq:exchange}
\Omega = \frac{e^2}{4 \pi \epsilon_0 \sqrt{m_1 m_2} \sqrt{\omega_1 \omega_2} d^3} \ .
\ee
This exchange means that energy can be removed from an initially hot ion by placing it close to a pre-cooled ``coolant'' ion for the exchange time $t_{\rm e} = \pi/(2 \Omega)$. This becomes useful when the coupling can be turned on and off without inducing excitation. One way to do this is to tune the two potential wells such that the ions come into resonance for a fixed time period, and subsequently detune them from each other. While this has been performed previously in the adiabatic regime \cite{Brown2011}, for the purposes of re-cooling ions in quantum computers it is desirable to increase the speed with which such operations are implemented. Thus we consider instead dynamic schemes.

\begin{figure}[!ht]
  \centering
    \includegraphics[width=0.5\textwidth]{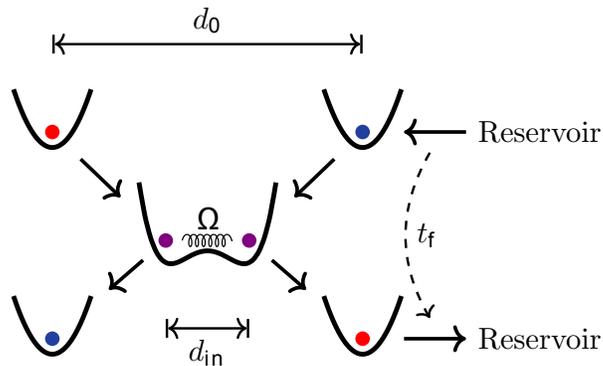}
  \caption{Schematic depiction of the overall cooling scheme. One ion is initially pre-cooled (blue circle) and interacts with a hot ion (red circle). After the cooling scheme, the ion energy has been transferred.}
\label{fig:scheme}
\end{figure}

Taking advantage of the strong $1/d^3$ scaling of the exchange frequency, we design an explicitly time-dependent Hamiltonian to transport two ions from an outer distance $d_{0}$ to an inner distance $d_{\rm in}$ and back out during a run-time $t_{\rm f}$, aiming to achieve a situation in which full energy transfer occurs. This scheme is illustrated in \fref{fig:scheme}. We explore the limits of the speed with which this can be carried out using trajectories designed with shortcuts to adiabaticity (STA) \cite{Guery2019}. Following a number of previous works \cite{Palmero2015, Ruschhaupt2012, Guery-Odelin2014, Lu2014, Zhang2015, Lu2015, Zhang2016, Lu2018, Levy2018}, we investigate the robustness of these methods to imperfections, and find solutions which are tolerant to these. Contrary to similar protocols in earlier work where strong approximations were made \cite{Lau14}, we consider a realistic double-well trapping potential including terms beyond the harmonic approximation, and assess the performance based on the full Hamiltonian of the system. The resulting schemes should allow robust cooling on timescales of 6 trap cycles, which is competitive with the operation speeds of high-fidelity multi-qubit gates \cite{Schaefer2018}.

The challenge of such a method is to design appropriate trajectories which do not add excitations during execution, such that the target ion may end up in the motional ground state. This can be realised by optimising the scheme with ground-state ions, disregarding the motional exchange at first. Thereafter, cooling can be achieved simply by finding the correct timing that leads to a complete exchange of energy.

This work is organised as follows. In \sref{sec:sec2}, we detail the physical constraints that the scheme needs to adhere to in order to be implementable in a real ion trap. Next, we obtain trajectories which we optimise to minimise the excitation of a pair of ground-state ions starting in separate potential wells. The theoretical techniques towards this end are presented in \sref{sec:sta}, whereas necessary numerical optimisation is carried out in \sref{sec:numerics}. In \sref{sec:robust_opt}, we add robustness against unwanted homogeneous electric fields. Having gained the ability to perform this basic protocol robustly and without additional excitations, we apply it in \sref{sec:cool} to find timings where two ions swap their energy. We then analyse the timing of these trajectories when varying the inner distance or the maximal quartic confinement in a given trap. 
In \sref{sec:resViol}, we consider the effect of a homogeneous field on the exchange of energy, and calculate the level of control of these fields that would be needed to successfully implement our scheme. Finally in \sref{sec:unequal}, the energy swapping protocol is generalised and optimised for two ions with unequal mass.

\section{Design constraints}
\label{sec:sec2}

The Hamiltonian for a system of two ions of the same mass $m$ is of the form
\be
H &=& \frac{p_1^2}{2m} + \frac{p_2^2}{2m} + V_\mathrm{tot}(x_1, x_2, t) \, \label{eq:Hamiltonian} \\
V_\mathrm{tot} &=& V_{\rm el}(x_1, t) + V_{\rm el}(x_2, t) + \frac{C_{\rm C}}{x_2-x_1}
\label{eq:symElPot}
\ee
where $\{x_i, p_i\}$ are the canonical position and momentum coordinates of the ions, $C_{\rm C} = e^2/(4 \pi \epsilon_0)$ and we assume that $x_2 > x_1$. In this equation the positions and momenta of the ions are time-dependent, although this is suppressed. We initially choose a symmetric quartic form for a double well potential \cite{Home2006}
\begin{equation}
\label{eq:quartPot}
V_\mathrm{el}(x,t) = \alpha (t)x^2 + \beta (t)x^4 \ .
\end{equation}
The assumption of symmetry will be relaxed in \sref{sec:unequal}, but provides a useful starting point to understand the methods that we use. The parameters $\{\alpha(t), \beta(t)\}$ describe the harmonic and quartic parts of the potential. As long as $\alpha(t)<0$ and $\beta(t)>0$, $V_{\mathrm{el}}$ is a double-well potential. This potential leads to an equilibrium ion separation $d$ which can be obtained from the equation 
\be
\label{eq:distance}
\beta d^5 + 2 \alpha d^3 - 2 C_{\rm C} = 0 
\ee
and normal mode frequencies for the excursions about the equilibrium positions of 
\be
\label{eq:nm_freq:-}
\Omega_{\rm -}^2 &=& \frac{1}{m}\left(2 \alpha + 3 \beta d^2\right) \\
\label{eq:nm_freq:+}
\Omega_{\rm +}^2 &=& \frac{1}{m}\left(2 \alpha + 3 \beta d^2 + \frac{4 C_{\rm C}}{d^3}\right) \; ,
\ee
where $\Omega_{\rm -}$ is associated with the in-phase motion (centre-of-mass mode) and $\Omega_{\rm +}$ with the out-of-phase motion (stretch mode).
Due to the symmetry of the potential, the potential curvatures at the ion positions are equal and given by 
\be
\label{eq:co-motFreq}
\omega_i^2 = \frac{1}{m}\left(2 \alpha + 3 \beta d^2 + \frac{2 C_{\rm C}}{d^3}\right) = \frac{1}{m}\left(2 \beta d^2 + \frac{4 C_{\rm C}}{d^3}\right) \ 
\ee
where $i = \{1,2\}$. We define the initial curvature as $\omega_0 = \omega_i(t=0)$.

The protocol should be designed to transport two ions from a separation of $d_0$ to a separation of $d_{\rm in}$ and back out again. For simplicity, we assume that the ions start and end the protocol at time $t_{\rm f}$ at the same separation $d(0) = d(t_{\rm f}) = d_0$. The distance $d_0$ should be chosen such that if no dynamical changes were made to the potential and the ions were simply kept at that distance during the protocol, the resulting exchange time $t_{\rm e}$ would be slow, which we take to mean that it is on the order of current cooling times and thus above 100 motional periods.

The ions are considered to achieve the distance of closest approach at the halfway point of the protocol $t_{\rm f}/2$, which we define as $d_{\rm in} = d(t_{\rm f}/2)$. We also choose to constrain our protocol by limiting the value of $\beta < \beta_{\rm max}$ , which requires the most demanding voltage settings for traps of the scale in use today \cite{Home2006}. To maintain ions in separate wells, which is the aim of our protocol, we can then see that $d_{\rm in}$ must be greater than
\be
\label{eq:critDist}
d_{\rm c} = \left( \frac{2C_{\rm C}}{\beta_{\rm max}}\right)^{1/5}  \ ,
\ee
which is obtained from \eref{eq:distance} with $\alpha = 0$. Attempting to aim for $d_{\rm in}<d_{\rm c}$ would involve combining the two ions into the same potential well, which involves taking the potential wells to conditions producing the lowest trap frequency for a given $\beta$. We make the assumption that this would be more experimentally challenging than keeping the ions in separated potentials.

From the considerations above, we can now set the parameters of the initial and final potential $\alpha_{\rm out} \equiv \alpha(0) = \alpha(\tf)$ and $\beta_{\rm out} \equiv \beta(0) = \beta(\tf)$ as well as the intermediate potential given by $\alpha_{\rm in} \equiv \alpha(\tf/2)$ and $\beta_{\rm in} \equiv \beta(\tf/2)$. To maintain two separate wells when the ions are closest, we take $\beta_{\rm in}$ to its maximum and thus $\beta_{\rm in} = \beta_{\rm max}$. From \eref{eq:distance} it then follows that
\be
\alpha_{\rm in} = \alpha(t_{\rm f}/2) = \frac{C_{\rm C}}{d_{\rm in}^3} - \frac{\beta_{\rm in} d_{\rm in}^2}{2} \ .
\ee

So far, the initial potential parameters $\alpha_{\rm out}, \beta_{\rm out}$ are only constrained by the selected outer distance $d_0$ and can otherwise be chosen freely. To fix them entirely, we choose the centre-of-mass frequency to be the same initially as at the half-way point of the protocol: $\Omega_{\rm -}(0) = \Omega_{\rm -}(\tf/2) = \Omega_{\rm -}(\tf)$. This then allows the determination of  $\alpha_{\rm out}, \beta_{\rm out}$ as well as the boundary values of the normal mode frequencies $\Omega_{0\pm} \equiv \Omega_{\pm}(0), \Omega_{\rm in\pm} \equiv \Omega_{\pm}(t_{\rm f}/2)$ by use of equations \eref{eq:distance}, \eref{eq:nm_freq:-} and \eref{eq:nm_freq:+}. Note that the initial curvature $\omega_0$ is then given by 

\be
\label{eq:omega0}
\omega_0^2 = \frac{1}{m}\left(2 \beta_{\rm out} d_0^2 + \frac{4 C_{\rm C}}{d_0^3}\right) \ .
\ee

In this way, the beginning, midpoint and final states of the protocol are dictated solely by these physical constraints, which consist of the desired initial and intermediate ion distances $d_0$ and $d_{\rm in}$, the maximal quartic confinement $\bmax$ that can be achieved in a given trap and the ion mass $m$.

The remaining task of designing suitable protocols is then to provide a transition between $\alpha_{\rm out}, \beta_{\rm out}$ and $\alpha_{\rm in}, \beta_{\rm in}$, and then back again. We make a working assumption that smooth transitions will provide the lowest excitations and use the symmetry of the problem to confine our search for protocols which are also symmetric in time around the point $t_{\rm f}/2$. To design protocols which achieve short execution times, we make use of so-called shortcut-to-adiabaticity techniques, which are described in the next section.

\section{Inverse engineering of a shortcut to adiabaticity}
\label{sec:sta}
The goal of STA techniques is to take a Hamiltonian $H(t)$ and design it in such a way that the populations in the instantaneous bases at $t=0$ and $t=t_{\rm f}$ are the same. This thus allows tasks to be executed in arbitrarily short times while yielding the same final result as an adiabatic evolution. STA methods have been proposed for many applications in trapped-ion QIP \cite{Palmero2015, Ruschhaupt2012, Lu2015, Torrontegui2011, Palmero2013, Palmero2014, Tobalina2017}. Of the various STA methods, we chose to use invariant-based inverse engineering as described in \cite{Chen2010}. 
This is an approach that involves first designing dynamical invariants $I(t)$ which commute with a general form of the system  Hamiltonian $H(t)$, and then deducing the explicit form of the latter from the resulting conditions. Dynamical invariants $I(t)$ are operators with constant expectation values
\begin{equation}
     \frac{\partial}{\partial t}\bra{\Psi}I(t)\ket{\Psi} = 0
\end{equation}
where $\ket{\Psi}$ are solutions to the Schrödinger equation for $H(t)$. For a Hamiltonian $H_\mathrm{HO}(t)$ with pure harmonic oscillator form, such invariants are known explicitly. This allows for the use of a result due to Lewis and Riesenfeld \cite{Lewis1969}, which states that if the Hamiltonian $H(t)$ and the corresponding invariants $I(t)$ are known, the individual solutions $\ket{\Psi}$ to the Schrödinger equation can be given as a superposition of eigenvectors $\ket{n;t}$ of $I(t)$:
\begin{equation}
\label{eq:superpos}
  \ket{\Psi} = \ket{\Psi(t)} = \sum_{n} c_{n} e^{i\alpha_{n}(t)}\ket{n;t}
\end{equation}
where the coefficients $c_{n}$ are constant and the phases $\alpha_{n}(t)$ are fully determined. This result is used in the invariant-engineering approach according to the following reasoning. If the invariant commutes with the Hamiltonian $\left([H_\mathrm{HO}(t_{\rm b}),I(t_{\rm b})] = 0\right)$ at boundary times $t_{\rm b} = \{0,t_{\rm f}\}$, they share an eigenbasis then. This means that the initial populations in the eigenbasis of $H_\mathrm{HO}(0)$ are also the populations of the invariant eigenvectors $\ket{n;0}$. Since Lewis-Riesenfeld theory tells us that the population numbers $c_{n}$ are constant, the system is still in the superposition $\sum_{n} c_{n} e^{i\alpha_{n}(t_{\rm f})}\ket{n;t_{\rm f}}$ at the final time, while the eigenvectors of the invariant $\ket{n;t_{\rm f}}$ have evolved. Since the Hamiltonian and the invariant commute again at $t_{\rm f}$, they again share an eigenbasis, meaning that the populations in the initial instantaneous basis of $H_\mathrm{HO}(0)$ have been carried over to the new basis of $H_\mathrm{HO}(t_{\rm f})$. This yields an evolution that recovers the initial populations at the final time. Note that the system may generally not follow the adiabatic evolution at intermediate times, but reaches the same final state nonetheless.

No invariant is known for the Hamiltonian in equation \eref{eq:Hamiltonian}. Thus in order to make use of STA methods, we first make an approximate transformation to a set of dynamical normal modes, for each of which the Hamiltonian takes the form of a harmonic oscillator at all times. The procedure that we follow is given in detail in \cite{Lizuain2017}. We make a second-order Taylor expansion of $H(t)$ about the equilibrium positions $x_1^{(0)}(t)$ and $x_2^{(0)}(t)$, and diagonalise the resulting mass-weighted Hessian matrix
\be
K_{ij} = \frac{1}{m}\frac{\partial^2V_{\rm tot}}{\partial x_i\partial x_j}\Big|_{\{x_1^{(0)},\; x_2^{(0)}\}}
\ee
of the potential $V_{\rm tot}$. For the potential defined in equation \eref{eq:quartPot}, the eigenvalues of $K$ are given by the squares of the dynamical normal mode frequencies $\Omega_\pm$ as defined from the relevant values of $\alpha(t), \beta(t)$. The eigenvectors are 
\be
v_\pm = \frac{1}{\sqrt{2}}\left(\begin{array}{c}
  1 \\
  \mp 1
 \end{array}\right)
\ee
and the corresponding normal mode coordinates
\be
\left(\begin{array}{c}
  X_- \\
  X_+
 \end{array}\right) &=&  \sqrt{\frac{m}{2}}\left(\begin{array}{c}
  x_2 + x_1\\
  (x_2 - x_1) - d
 \end{array}\right) \\
 \left(\begin{array}{c}
  P_-\\
  P_+
\end{array}\right) &=&  \sqrt{\frac{1}{2m}}\left(\begin{array}{c}
  p_2 + p_1\\
  p_2 - p_1 + m\dot{d}
\end{array}\right)
\ee
include the dynamic component $\dot{d}$. In terms of these coordinates the Hamiltonian can be written as
\be
\label{eq:equalHam}
H &\approx& H_\mathrm{2HO}= H_\mathrm{HO}^{(+)} + H_\mathrm{HO}^{(-)}\nonumber\\ &=&\underbrace{\frac{{P}_+^2}{2} + \frac{1}{2}\Omega_+^2\left({X}_+ + \sqrt{\frac{m}{2}}\frac{\ddot{d}}{\Omega_+^2}\right)^2}_{H_\mathrm{HO}^{(+)}} +
 \underbrace{\frac{{P}_-^2}{2} + \frac{1}{2}\Omega_-^2{X}_-^2}_{H_\mathrm{HO}^{(-)}}
\ee
which also includes the dynamic component $\ddot{d}$. Inserting these Hamiltonians into the results of \ref{app:LRtheory} (equations \eref{eq:generalInvariant} through \eref{eq:genInstEnergies}), we find the corresponding Lewis-Riesenfeld invariants for each dynamical normal mode to be
\begin{equation}
\label{eq:equalPMInvariant}
I^{(\pm)} = \frac{1}{2}\left[\rho_\pm(P_\pm - \dot{q}_\pm) - \dot{\rho}_\pm(X_\pm - q_\pm)\right]^2 +
 \frac{1}{2}\Omega_{0\pm}^2\left(\frac{X_\pm - q_\pm}{\rho_\pm}\right)^2 \ ,
\end{equation} 
where the auxiliary functions $\rho_\pm$ and $q_\pm$ are defined by
\be
\label{eq:equalAuxODE}
    \ddot{\rho}_\pm + \Omega_\pm^2&\rho_\pm =& \frac{\Omega_{0\pm}^2}{\rho_\pm^3}\label{eq:equalAuxODE:rho}\\
    \ddot{q}_+ + \Omega_+^2&q_+ =& -\sqrt{\frac{m}{2}}\ddot{d}\label{eq:equalAuxODE:x}\\
    &q_- =&0 \ . \label{eq:equalAuxODE:x_min} 
\ee
The mode energies are given by 
\be
\label{eq:specInstEnergies+} 
\fl E^{(+)}_{n} &= \frac{\hbar(2n+1)}{4\Omega_{0+}}\left(\dot{\rho}_+^2 + \Omega_+^2\rho_+^2 +\frac{\Omega_{0+}^2}{\rho_+^2}\right)
+ \underbrace{\frac{1}{2}\dot{q}_+^2 + \frac{1}{2}\Omega_+^2\left(q_+ + \sqrt{\frac{m}{2}}\frac{\ddot{d}}{\Omega_+^2}\right)^2}_{\eqdef E^{(+)}_{\rm q}} \\
\fl E^{(-)}_{n} &= \frac{\hbar(2n+1)}{4\Omega_{0-}}\left(\dot{\rho}_-^2 + \Omega_-^2\rho_-^2+\frac{\Omega_{0-}^2}{\rho_-^2}\right) \ .\label{eq:specInstEnergies-} 
\ee
Note that we have defined here the part $E^{(+)}_{\rm q}$ of the stretch-mode energy that pertains to the auxiliary function $q_+$, such that we may minimise its influence later on.

The physical meaning of the auxiliary variables $q_\pm$ can be understood as the normal mode centres, while $\rho_\pm$ correspond to the effective curvatures of the oscillator potential in the transformed co-ordinates. As an example, in the case of transport of one ion in a constant trapping well the function $\rho$ can be set to 1, as the ion experiences the same potential curvature at all times. In the case we are currently considering, $q_-$ is zero at all times due to the spatial symmetry of the potential.

In order to make use of the invariants, we must first find a parametrisation which satisfies the commutation relation $\left[H_{\rm HO}^{(\pm)}, I^{(\pm)}\right]$ at the boundary times $t_{\rm b} = \{0, t_{\rm f}\}$. The commutation can be ensured by setting the boundary conditions
\be
\label{eq:BC}
    \rho_\pm (t_{\rm b}) &=& 1\  \\
\dot{\rho}_\pm(t_{\rm b}) &=& \ddot{\rho}_\pm(t_{\rm b}) = \rho^{(3)}_\pm(t_{\rm b}) = \rho^{(4)}_\pm(t_{\rm b}) = 0\label{eq:BC:rho}\\
    q_+(t_{\rm b}) &=& \dot{q}_+(t_{\rm b}) = \ddot{q}_+(t_{\rm b}) = 0\label{eq:BC:x}
\ee
on the auxiliary functions.
The conditions on the zeroth and first derivatives arise from the Lewis-Riesenfeld theory (see \ref{app:LRtheory} for further details). In addition, we constrain the first two derivatives of the distance $\dot{d}(t_{\rm b})$ and $\ddot{d}(t_{\rm b})$ to zero, so that the scheme starts and ends with stationary ions and to minimise the energy. This together with the auxiliary equations \eref{eq:equalAuxODE:rho} and \eref{eq:equalAuxODE:x} leads to the remaining conditions.

To fulfil the physical constraints at the midpoint of the protocol, the normal mode frequencies need to reach $\Omega_{\rm in \pm}$ at $t_{\rm f}/2$, which is achieved approximately by setting 

\be
\label{eq:BC_half}
 \rho_{\rm in\pm} \equiv \rho_\pm (t_{\rm f}/2) =  \sqrt{\frac{\Omega_{0\pm}}{\Omega_{\rm in\pm}}}\ .
\ee
This can be verified by inserting the condition into equation \eref{eq:equalAuxODE} and neglecting the term in $\ddot{\rho}_\pm$, which is justified as long as the protocol takes several motional cycles.

Any choice of $\{\rho_\pm, q_\pm\}$ satisfying the boundary conditions above leads to a shortcut to adiabaticity with respect to the Hamiltonian $H_{\rm 2HO}$, leaving flexibility to optimise the scheme for various purposes, such as cancellation of residual excitations or robustness to experimental imperfections. However choosing $\{\rho_\pm, q_\pm\}$ to simultaneously fulfil the ODEs in \eref{eq:equalAuxODE:rho} - \eref{eq:BC:x} is hard. Therefore we follow \cite{Palmero2015} and design a general form for $\rho_\pm$ which satisfies only \eref{eq:equalAuxODE:rho}, \eref{eq:BC} and \eref{eq:BC:rho}, but contains additional free parameters. We then perform a numerical search in the free parameter space in order to obtain solutions which are as close as possible to satisfying the additional constraints in \eref{eq:BC:x}. One way of doing so is to minimise the part $E^{(+)}_{\rm q}(t_{\rm f})$ of the mode energy in \eref{eq:specInstEnergies+} pertaining to the auxiliary $q_{\rm +}$, thus finding parameters that come close to fulfilling $q_+(t_{\rm f}) = \dot{q}_+(t_{\rm f}) =  \ddot{q}_+(t_{\rm f}) = 0$.

We expect that a smoothly varying function for $\rho_\pm(t)$ would be the most satisfactory experimentally. For this reason, we use a polynomial interpolation function for $\rho_\pm$, which contains only even orders due to the chosen symmetry of the protocol. The simplest polynomial satisfying the boundary conditions for the centre-of-mass mode is 
\be
\rho_- &=&  1
\label{eq:rho_min}
\ee
due to having chosen $\Omega_{0-}= \Omega_{\rm in-}$ earlier.

There are 11 constraints on $\rho_+(t)$ given by \eref{eq:BC},\eref{eq:BC:rho} and \eref{eq:BC_half}. In addition to that, we choose $\rho_+(t)$ to be symmetric about $t_{\rm f}/2$. Thus setting $\rho_+(t)$ to be a polynomial of order 14 leaves two free parameters for numerical optimisation. The resulting $\rho_+(t)$ fulfilling the constraints is 
\be
\rho_+(s) &=&  \rho_{\rm in+} + \frac{A}{2} \left(s-\frac{1}{2}\right)^2 \nonumber \\
&+& \frac{1}{1024}\left(245760 - 10240A - 245760\rho_{\rm in+}-B\right) \left(s-\frac{1}{2}\right)^4\nonumber \\
&-& \frac{5}{256}\left(131072 - 4096A - 131072\rho_{\rm in+}-B\right) \left(s-\frac{1}{2}\right)^6\nonumber \\
&+& \frac{5}{32}\left(73728 - 2048A - 73728\rho_{\rm in+}-B\right) \left(s-\frac{1}{2}\right)^8\nonumber \\
&-& \frac{1}{8}\left(196608 - 5120A + 196608\rho_{\rm in+}-5B\right) \left(s-\frac{1}{2}\right)^{10}\nonumber \\
&+& \frac{1}{4}\left(81920 - 2048A - 81920\rho_{\rm in+}-5B\right) \left(s-\frac{1}{2}\right)^{12}\nonumber \\
&+& B \left(s-\frac{1}{2}\right)^{14}.
\label{eq:rho_pl}
\ee
with the normalised time $s=t/t_{\rm f}$ and the two free parameters $A$ and $B$. The former constrains the curvature of $\rho_+$ at $s=1/2$, while the latter is the co-efficient of the 14\textsuperscript{th} order term.

A protocol can now be fully defined in the following way: First the physical constraints $d_{\rm out}$, $d_{\rm in}$, $\bmax$ and $m$ are defined, from which the boundary parameters $\alpha_{\rm out}, \beta_{\rm out}$ and $\alpha_{\rm in}, \beta_{\rm in}$ as well as the boundary mode frequencies $\Omega_{0\pm}$, $\Omega_{\rm in \pm }$ can be calculated as described in \sref{sec:sec2}. From this follows $\rho_{\rm in+}$ and the ansatz \eref{eq:rho_pl} is completed by a choice of the free parameters $A$ and $B$.
Finally, we want to obtain the time dependence of the potential parameters $\alpha(t)$ and $\beta(t)$. We first solve \eref{eq:equalAuxODE:rho} for $\Omega_\pm$ and observe that $\Omega_- = \Omega_{0-}$. These together with \eref{eq:distance}, \eref{eq:nm_freq:-} and \eref{eq:nm_freq:+} then yield the desired functions.

In the following section, we utilise the free parameter $A$ to find optimised trajectories that do not create residual excitations when being executed on two ground-state ions. This is a prerequisite for achieving  motional energy swapping. Furthermore, by using the parameter $B$ as well, the protocol can be made robust to common experimental imperfections, for which the necessary steps are worked out in section \ref{sec:robust_opt}.

\section{Numerical optimisation of the shortcut for low residual excitations}
\label{sec:numerics}

To achieve the goal of designing trajectories which achieve minimal excitation when transporting two ground-state ions from a separation of $d_0$ to one of $d_{\rm in}$ and back again, we now optimise the ansatz $\rho_+$. Two obstacles stand in the way of analytically designing a perfect shortcut to adiabaticity for the given double-well potential. First of all, the ansatz for the auxiliary function $\rho_+$ does not a-priori guarantee that $q_+$ also fulfils the boundary conditions. Secondly, the shortcut was designed for the harmonically approximated Hamiltonian $H_{\rm2HO}$ and can therefore not result in an excitation-free scheme for the full Hamiltonian.

This motivates the comparison of two different approaches, the first being the simple completion of the shortcut to adiabaticity by finding the value of the parameter $A$ (while keeping $B=0$) that yields a minimum of the part $E^{(+)}_{\rm q}$ of the stretch-mode energy. In the second method we choose instead to optimise the single free parameter by directly minimising the exact final excitation of the full Hamiltonian. Previous work \cite{Palmero2015} attempted to reduce the effect of the discrepancy between $H$ and $H_{\rm2HO}$ by including higher order corrections to the mode energies and optimising additional free parameters. The method presented here is numerically more costly, but we found that it was not a significant problem for the trajectories which we considered, while yielding a better performance.

Note again that, once an optimised shortcut is found, we will aim in \sref{sec:cool} to find run-times $t_{\rm f}$ that yield a complete energy swap. This means that we now need to optimise the shortcut ansatz separately for a range of run-times, for each of which different values for the parameter $A$ may be found.

\subsection{Numerical prerequisites}

To study the effectiveness of both optimisation approaches, we need to compute the final energy of the ions after a given protocol. Numerical integration of the classical equations of motion given by the full Hamiltonian in \eref{eq:Hamiltonian} is used to calculate this. However, the total energy increase cannot be divided between the individual ions, as they are coupled by the Coulomb interaction. We therefore expand the Coulomb potential in \eref{eq:Hamiltonian} to first order in the displacement of the ions from the equilibrium positions and calculate the energy increase of one of the ions at time $\tf$ with respect to the equilibrium energy as
\be
E_{\rm ex,i}(\tf) &=&\frac{p_i^2(\tf)}{2m} +  \left[V_{\rm el}\left(x_i(\tf),\tf\right) - V_{\rm el}\left(x_i^{(0)}(\tf),\tf\right)\right]\nonumber \\
&-& (-1)^i \frac{C_{\rm C}}{d^2(\tf)}\Big(x_i(\tf) - x_i^{(0)}(\tf)\Big)\ ,
\label{eq:Eex}
\ee
where $i = 1,2$ denotes the ion. The positions $x_i(t)$ and momenta $p_i(t)$ of the ions are found by integrating the equations of motion. Note that when using the the symmetric electrical potential \eref{eq:symElPot}, both ions have the same energy increase $E_{\rm ex,1}=E_{\rm ex,2}$. This will not hold in later sections when we consider asymmetric potentials.

In this section, the ions are assumed to be in their ground state at $t=0$, which is implemented by setting their initial momentum to zero and placing them at the positions of the trapping potential minima. The physical constraints are shown in table \ref{table:target-values} and reflect a realistic experiment. The achievable $\beta_{\rm max}$ (and with it, the critical distance $d_{\rm c}$) is that of the Sandia HOA2 surface trap \cite{Maunz2016} recently used in several research groups \cite{Stephenson2019, Crain2019, Tabakov2018}, including our own. It is obtained from simulations of the trap together with the assumption that potentials with a voltage range of $\pm 10 {\rm V}$ are supplied to the electrodes.
\begin{table}[ht]
\caption{\label{table:target-values}List of the default set of physical constraints used for numerical examples.}
\begin{indented}
\item[]\begin{tabular}{@{}lllll}
\br
$\beta_{\rm max} \left(\si{\newton\per\meter\cubed}\right)$ & $d_{\rm c} \left(\si{\micro\meter}\right)$ & $d_0 \left(\si{\micro\meter}\right)$ & $d_{\rm in} \left(\si{\micro\meter}\right)$ & ${\rm m} \left(\si{\amu}\right)$\\
\mr
$ \num[{scientific-notation = true}]{0.85e-3}$ & $14.0$ & $5d_{\rm c} = {70.1}$  & $1.1d_{\rm c} = {15.4}$  & {39.96 $\left(^{40}{\rm Ca}^+\right)$}\\
\br
\end{tabular}
\end{indented}
\end{table}

The inner distance $d_{\rm in}$ was chosen slightly above the critical distance $d_{\rm c}$, such that the ions never get merged into a single well. The outer distance $d_0$ was chosen such that the exchange interaction is slow compared to the targeted protocol run-time. At the outer distance given in table \ref{table:target-values}, the exchange time $t_{\rm e}$ is about $\SI{442}{\micro\second}$. As the initial curvature $\omega_0$ is about \SI{0.45}{\mega\hertz}, this corresponds to 200 trap periods.

\subsection{Minimisation of residual excitations}

\begin{figure}[ht]
    \centering
        \subfloat{\includegraphics[width=0.5\columnwidth]{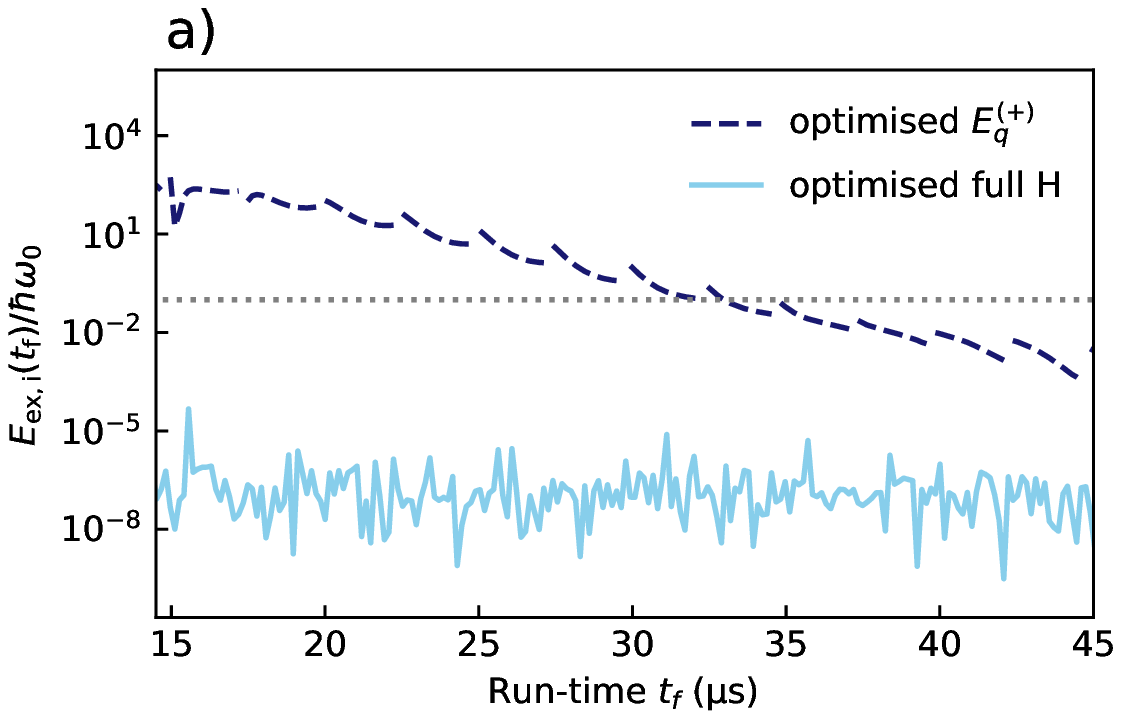}}
        \subfloat{\includegraphics[width=0.5\columnwidth]{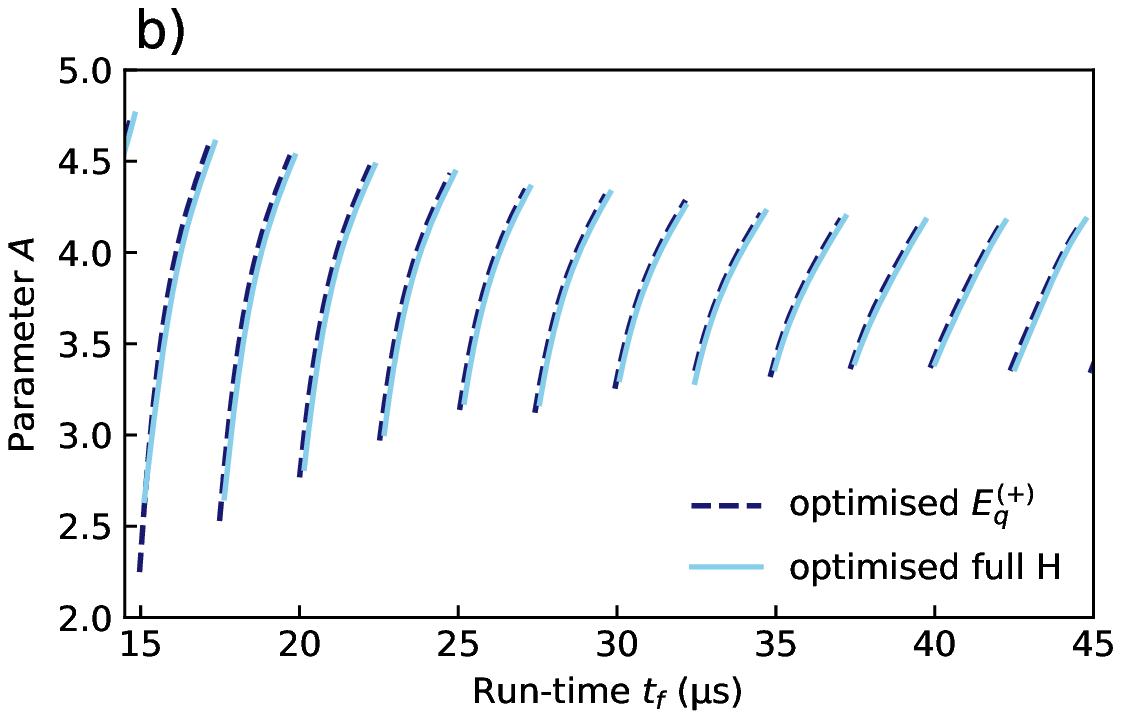}\label{fig:OneParam17Comp:b}}
    \caption{Comparison of optimisation methods to obtain a shortcut to adiabaticity. Shown are (a) the resulting final excitations $E_{\rm ex,i}(\tf)$ after minimising $E^{(+)}_{\rm q}$ (dashed lines) and the full Hamiltonian (solid lines) and (b) the corresponding optimal values of the free parameter $A$. The physical design constraints used for the trajectories are given in \tref{table:target-values}. The numerical optimisation was run for each run-time separately. The excitation $E_{\rm ex,i}$ of one of the ions is normalised by the energy $\hbar\omega_0$ of one phonon, where $\omega_0$ is the initial potential curvature. The dotted line in a) corresponds to final excitations of 0.1 quanta and marks the energy level below which the scheme can in principle yield cooling of one of the ions to better than 0.1 quanta, which we consider the target. Note that $E_{\rm ex,1} = E_{\rm ex,2}$ holds here due to the symmetry of the problem.}
    \label{fig:OneParam17Comp}
\end{figure}

The results of the two approaches are compared in \fref{fig:OneParam17Comp}. For each run-time shown, the free parameter $A$ has been optimised using the two methods and the exact final excitation $E_{\rm ex,i}(\tf)$ is plotted. Numerically, the optimisation is performed with the Nelder-Mead algorithm provided by the NumPy package in Python. When $E^{(+)}_{\rm q}$ is used as a minimisation target, $E_{\rm ex,i}(\tf)$ is not zero as it should be if a shortcut has indeed been found, instead it increases exponentially with shorter run-times, consistent with the behaviour reported in \cite{Palmero2015}. This is due to break down of the harmonic approximation used in constructing the dynamical normal modes. On the other hand, we are able to reduce the excitation to the level of numerical tolerance when optimising the full Hamiltonian directly and without increasing the number of free parameters used.

In \fref{fig:OneParam17Comp:b}, the optimised parameter $A$ depending on the run-time is depicted. The two optimisation approaches yield similar values, yet the difference in final excitations grows to more than ten orders of magnitude at a run-time of \SI{15}{\micro\second}. We have thus demonstrated a way to numerically construct a shortcut to adiabaticity and perform the proposed protocol such that no additional excitations are added.

\subsection{Specific example trajectory}

\begin{figure}[!ht]
  \centering
    \includegraphics[width=1\textwidth]{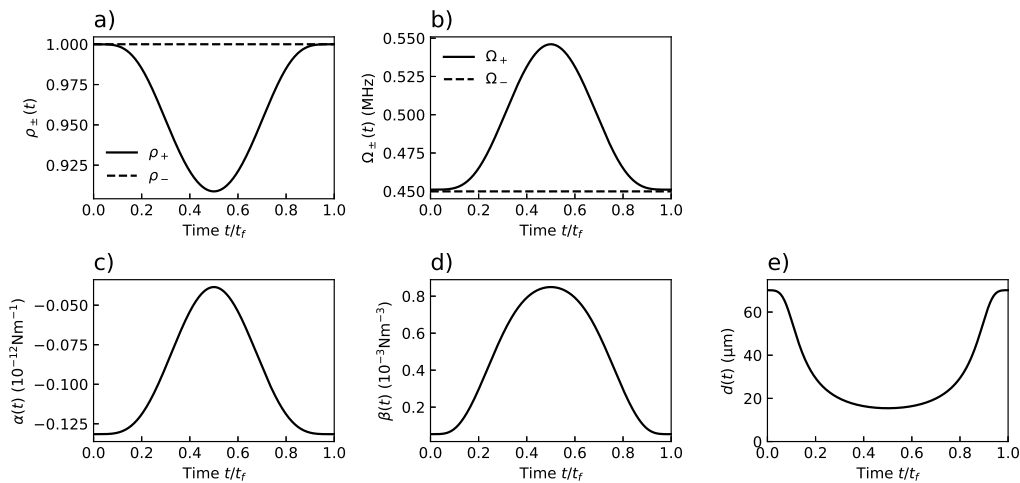}
  \caption{Time evolution of an optimised protocol. The optimal value of $A$ found with the full Hamiltonian method at $t_{\rm f}= \SI{30}{\micro\second}$ is used, as presented in \fref{fig:OneParam17Comp}. The protocol is designed to meet the constraints in \tref{table:target-values}. Shown are (a) the auxiliary functions $\rho_{\rm \pm}(t)$, (b) the normal mode frequencies $\Omega_{\rm \pm}(t)$, (c) the potential constituents $\alpha(t)$ and (d) $\beta(t)$ and (d) the resulting distance trajectory $d(t)$.}
\label{fig:opt-overview}
\end{figure}

For illustration, the parameter functions of a protocol optimised in this way is shown in \fref{fig:opt-overview}. We pick the trajectory found by using the full Hamiltonian as an optimisation target at a run-time of $t_{\rm f}=\SI{30}{\micro\second}$. Shown is the time evolution of the auxiliary functions $\rho_{\rm \pm}(t)$, the normal mode frequencies $\Omega_{\rm \pm}(t)$, the potential constituents $\alpha(t)$ and $\beta(t)$ as well as the resulting distance trajectory $d(t)$. Note that the centre-of-mass mode frequency $\Omega_-$ is constant as designed and that the physical constraints specified in table \ref{table:target-values} are met. Despite the high-order ansatz, the optimised trajectories are smooth.

We could now go on to explore how to achieve cooling. However, the protocols found in this section are susceptible to experimental imperfections. We therefore defer the analysis of cooling solutions to section \ref{sec:cool} and first address experimental robustness in the following section.

\section{Robustness optimisation of the shortcut}
\label{sec:robust_opt}

A commonly encountered experimental imperfection in trapped-ion QIP is an additional homogeneous electric field \cite{Kaufmann2014, Ruster2014}, which could arise due to a number of sources, such as stray charges in the trap or inaccuracies in the applied trap voltages. It can also be regarded as the first order expansion of a general perturbation potential. By including this homogeneous field in the derivation of the STA, we find modified expressions for the mode energies. These can be leveraged to achieve optimised robustness towards this type of experimental imperfection.

\subsection{Robustness condition}
\label{sec:robust}

A given double-well potential is altered by such a field to
\begin{equation}
\label{eq:pertPotEqual}
\tilde{V}_{\rm el}(t,\tilde{x}) = \gamma \tilde{x} + \alpha (t)\tilde{x}^2 + \beta (t)\tilde{x}^4,
\end{equation}
where a linear term $\gamma$ was added. Variables of the perturbed system are indicated with a tilde in this section. 
The equilibrium positions $\tilde{x}_i^{(0)}$ of the ions are not symmetric anymore and we introduce the centre-of-mass shift $\tilde{s}$ to denote $\tilde{x}_1^{(0)} = \tilde{s} -\tilde{d}/2$ and $\tilde{x}_2^{(0)} = \tilde{s} + \tilde{d}/2$. The ion distance is given by $\tilde{d}$, which may differ from the unperturbed $d$. In what follows we assume that the additional field is weak, such that the shift $\tilde{s}$ is much smaller than the equilibrium distance, allowing us to include it only to first order.
Under this assumption, we find that $\tilde{d} \approx d$ and
\begin{equation}
\label{eq:sShift}
\tilde{s} \approx -\frac{\gamma}{2\alpha + 3\beta d^2} = -\frac{\gamma}{m\Omega_-^2}.
\end{equation}
To describe the strength of the imperfection in a way that allows a dimensionless small parameter expansion, we define the perturbation parameter
\begin{equation}
\label{eq:lambdaParameter}
\eta \equiv \frac{-\tilde{s}}{d_{\rm in}} = \frac{\gamma}{m\Omega_-^2d_{\rm in}},
\end{equation}
which turns the linear approximation condition $|\tilde{s}| \ll \tilde{d}$ into $\eta = |\tilde{s}|/d_{\rm in} \ll 1$.
Following \cite{Lizuain2017}, the normal mode eigenfrequencies stay the same to first order: $\tilde{\Omega}_\pm = \Omega_\pm$, but the corresponding eigenvectors become
\be
\tilde{v}_\pm &\approx \left(\begin{array}{c}
  1 \mp 2\Delta \\
  \mp 1
 \end{array}\right),
\ee
where the first entry has been shifted by $\Delta = 3C_{\rm C}^{-1}\beta d^4 \tilde{s}$.  This leads to the normal mode position/momentum coordinates becoming
\be
\left(\begin{array}{c}
  \tilde{X}_- \\
  \tilde{X}_+
 \end{array}\right) &\approx&  \tilde{A}(t)\left(\begin{array}{c}
  \tilde{x}_1 - (\tilde{s}-\frac{d}{2})\\
  \tilde{x}_2 - (\tilde{s}+\frac{d}{2})
 \end{array}\right)\\
 \left(\begin{array}{c}
   \tilde{P}_-\\
  \tilde{P}_+
\end{array}\right) &\approx&  (\tilde{A}^T)^{-1}(t)\left(\begin{array}{c}
  \tilde{p}_1\\
  \tilde{p}_2
 \end{array}\right) + \sqrt{\frac{m}{2^3}}\left(\begin{array}{c}
  \dot{\tilde{d}}\Delta\\
  \dot{\tilde{d}}
 \end{array}\right)
\ee
with the coordinate change matrix
\be
\tilde{A}(t) \approx \sqrt{\frac{m}{2}}\left(\begin{array}{c c}
  1-\Delta & 1+\Delta\\
  -(1+\Delta) & 1-\Delta
 \end{array}\right) \ .
\ee
The Hamiltonian in these coordinates is given by 
\be 
\label{eq:pert2HO}
\fl{\tilde{H}}_{\rm 2HO} &=& {\tilde{H}}^{(+)}_{\rm HO} + \tilde{H}^{(-)}_{\rm HO} + H_{\rm c} \\ \fl&=& \underbrace{\frac{{\tilde{P}^{2}_+}}{2} + \frac{1}{2}\tilde{\Omega}_+^2\left(\tilde{X}_+ + \sqrt{\frac{m}{2}}\frac{\ddot{\tilde{d}}}{\tilde{\Omega}_+^2}\right)^2}_{{\tilde{H}}^{(+)}_{\rm HO}} + \underbrace{\frac{{\tilde{P}^{2}_-}}{2} + \frac{1}{2}\tilde{\Omega}_-^2\left(\tilde{X}_- + \sqrt{\frac{m}{2}}\frac{\ddot{\tilde{d}}\Delta}{\tilde{\Omega}_-^2}\right)^2}_{\tilde{H}^{(-)}_{\rm HO}}\nonumber\\ \fl&&+ \underbrace{\dot{\Delta}\left(\tilde{X}_+\tilde{P}_- - \tilde{X}_-\tilde{P}_+\right)}_{H_{\rm c}} \ .\nonumber
\ee

From this form, and ignoring the term in $\dot{\Delta}$ for the moment, we observe that the stretch-mode Hamiltonian has retained its form in the new variables, implying that the definitions of the auxiliary functions of the stretch-mode $\rho_+$ and $q_+$ in \eref{eq:equalAuxODE:rho} and \eref{eq:equalAuxODE:x} remain unchanged as well as the mode energy $E^{(+)}_{n}$ in \eref{eq:specInstEnergies+}. Since the COM-mode part has gained a term in $\Delta$, \eref{eq:specInstEnergies-} and \eref{eq:equalAuxODE:x_min} become invalid for the perturbed system. By applying the results of \ref{app:LRtheory}, the definition of $q_-$ changes instead to be
\be 
\ddot{\tilde{q}}_- + \tilde{\Omega}_-^2\tilde{q}_- = -\sqrt{\frac{m}{2}}\ddot{\tilde{d}}\Delta
\ee
and the perturbed mode energy becomes 
\be
\fl\tilde{E}^{(-)}_{n} = \frac{\hbar(2n+1)}{4\tilde{\Omega}_{0-}}\left(\dot{\tilde{\rho}}_-^2 + \tilde{\Omega}_-^2\tilde{\rho}_-^2+\frac{\tilde{\Omega}_{0-}^2}{\tilde{\rho}_-^2}\right) + \overbrace{\frac{1}{2}\dot{\tilde{q}}_-^2 + \frac{1}{2}\tilde{\Omega}_-^2\left(\tilde{q}_- + \sqrt{\frac{m}{2}}\frac{\ddot{\tilde{d}}\Delta}{\tilde{\Omega}_-^2}\right)^2}^{\tilde{E}^{(-)}_q} \ . \label{eq:pertSpecInstEnergies-} 
\ee
Hence there are now further boundary conditions to be fulfilled in addition to those in \eref{eq:BC}, namely 

\be 
\tilde{q}_-(t_{\rm b}) &=& \dot{\tilde{q}}_-(t_{\rm b}) = \ddot{\tilde{q}}_-(t_{\rm b}) = 0\label{eq:BC:x_min} \ . 
\ee
By adding these constraints to the optimisation of the free parameters $A$ and $B$ presented in section \ref{sec:numerics}, we are able to achieve optimal robustness with respect to a linear potential perturbation. Analogous to the procedure in \sref{sec:numerics}, this is most easily achieved by minimising the part $\tilde{E}^{(-)}_q$ of the final energy $\tilde{E}^{(-)}_{n}$ in addition to minimising $E^{(+)}_{q}$ as before.

Equation \eref{eq:pert2HO} also contains a mode-coupling term $H_{\rm c}$, which could only be decoupled in a further transformation if $\dot{\Delta}$ was time-independent \cite{Lizuain2019}. Thus we ignore it and accept that any protocol optimised in this way will display some degree of mode coupling in the presence of such a homogeneous field. We see numerically that this does not produce a great problem for the situations considered here.

\subsection{Frequency mismatch}
\label{ssec:mismatch}

Here we should point out that an additional homogeneous field does not only cause excess excitations, but also introduces a mismatch between the potential curvatures at the two ions due to the quartic term in the potential. From the perturbed normal mode calculation, one can calculate the shifted curvatures $\tilde{\omega}_{i}$ due to a perturbation $\eta$ as

\be
\label{eq:freq_mis}
    m\tilde{\omega}_{i}^2 &\approx 2\beta d^2 + \frac{4C_{\rm C}}{d^3} \mp 12\beta d \tilde{s} = m\omega_{i}^2 \mp 12\beta d \tilde{s},
\ee
making the mismatch $\delta\omega = \tilde{\omega}_{2} - \tilde{\omega}_{1} \approx 12\beta d \tilde{s}/\left(m \omega_i\right)$ to first order. 

This is problematic since the motional exchange \eref{eq:exchange} is a resonant effect with respect to the potential curvatures, meaning that such a homogeneous field can suppress the desired energy exchange. Note that no protocol optimisation can mitigate this off-resonance  effect. We deduce the required level of experimental accuracy that this imposes on the protocol in \sref{sec:resViol}.

\subsection{Numerical optimisation for optimal robustness and low residual excitations}
\label{sec:robust_numeric}

Since a real experiment will never be able to implement the protocols found in \sref{sec:numerics} without error, we incorporate the robustness results of section \ref{sec:robust} into the numerical optimisation. To make sure that all the boundary conditions, including the ones on $\tilde{q}_-$ and $q_+$ are fulfilled, we minimise the parts of the mode energies related to these auxiliary functions by choosing the cost function 

\be
\label{eq:cost-app}
   {\rm Approximate \: cost} = E_{q}^{(+)}(t_{\rm f}) + \tilde{E}_{q}^{(-)}\left(t_{\rm f},\eta=0.015\right)
\ee
This ensures that a shortcut is indeed constructed (in the harmonic approximation that yielded \eref{eq:equalHam}) and the effects of a perturbative field are minimised. Since $\tilde{E}_{q}^{(-)}$ depends on the strength of the perturbation $\eta$, we choose a value of $\eta = 0.015$ which, at the design parameters in table \ref{table:target-values}, corresponds to an electrical field value of about \SI{1.0}{\volt\per\meter}.

In \sref{sec:numerics}, we obtained protocols with lower final excitations by optimising the energy of the full Hamiltonian instead of the energy $E_{q}^{(+)}$ obtained by a harmonic approximation. To see if the same is true when including robustness in the optimisation, we define a further cost function analogous to \eref{eq:cost-app}, but only consisting of excitations obtained by integrating the equations of motion of the full Hamiltonian:

\be
\label{eq:cost-ex}
   {\rm Exact \: cost} = \sum_{i=1}^2 E_{\rm ex,i}(t_{\rm f},\eta=0) + E_{\rm ex,i}(t_{\rm f}, \eta=0.015) \ .
\ee
To calculate the final excitation $E_{\rm ex,i}(t_{\rm f}, \eta)$ of ion $i$, the perturbed potential $\tilde{V}_{\rm el}$ is inserted into the full Hamiltonian \eref{eq:Hamiltonian}, which we will now denote with $H_\eta$. The initial conditions are defined by placing the ions at rest at the equilibrium positions of this perturbed potential. The energy is then calculated in the same way as in \eref{eq:Eex}. Note that $E_{\rm ex,1} = E_{\rm ex,2}$ does not hold anymore for non-zero $\eta$.

For both cost functions, we use a Nelder-Mead algorithm to find the minimum, this time by varying both free parameters $A$ and $B$. As before in \sref{sec:numerics}, we re-perform this optimisation for each considered run-time $t_{\rm f}$.

\begin{figure}[!ht]
  \centering
    \includegraphics[width=1\textwidth]{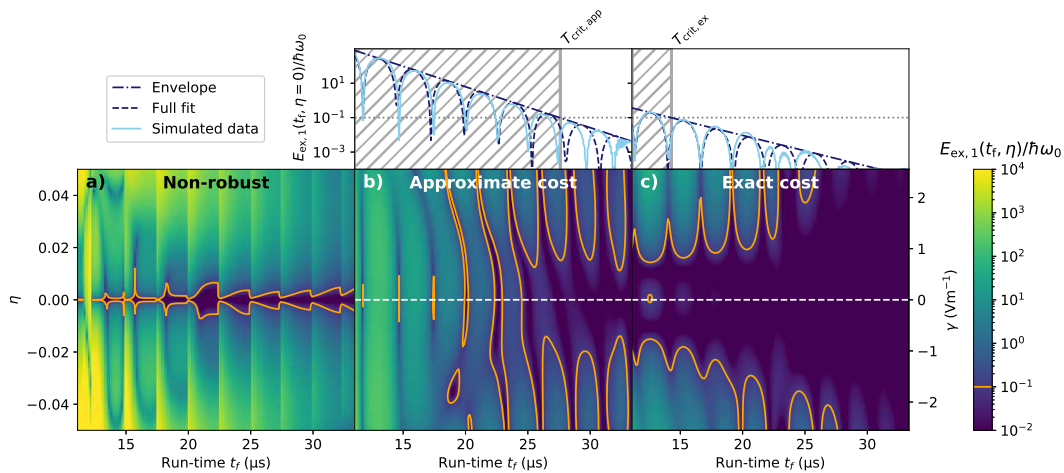}
  \caption{Comparison of protocol optimisation methods. (a) Effect of a perturbation on the non-robust trajectories shown in \fref{fig:OneParam17Comp}. (b) Robustness to perturbations of the trajectories obtained with the approximate cost function. (c) Robustness to perturbations of the trajectories obtained with the exact cost function. The excitation energies are normalised by the single phonon energy $\hbar \omega_0$, where $\omega_0$ is about $2\pi\cdot\left(\SI{0.45}{\mega\hertz}\right)$ for these physical constraints. The orange contour lines denote an excitation level of $\bar{n}=0.1$. In plot a), we have reused the optimised values of $A$ from \fref{fig:OneParam17Comp} and plot the final excitations $E_{\rm ex,1}(t_{\rm f}, \eta)$. To obtain plots b) and c), the free parameters $A$ and $B$ are optimised anew for each run-time, using the two cost functions \eref{eq:cost-app} and \eref{eq:cost-ex}. The final excitation $E_{\rm ex,1}(t_{\rm f},\eta)$ of one of the ions is calculated on a grid of run-times and perturbation parameters, assuming the ions to be in their ground state initially. Note that plotting $E_{\rm ex,2}$ results in the same data, but mirrored along the line $\eta = 0$. The physical constraints for all shown schemes are those from \tref{table:target-values}. The perturbation is also shown in terms of the homogeneous field $\gamma$, which is obtained from $\eta$ using \eref{eq:lambdaParameter}.  The insets above b) and c) show a cut through the data (indicated by the white dashed lines), corresponding to the unperturbed final excitations $E_{\rm ex,1}(t_{\rm f}, \eta=0)$ together with a fit to that data and an envelope function. The dotted line in the insets corresponds to final excitations of 0.1 quanta and the hatched area marks the range of run-times where the envelope indicates final excitations above this level.}
\label{fig:robust}
\end{figure}

The resulting performance of the two methods is displayed in \fref{fig:robust}. For each run-time $t_{\rm f}$, the two optimisation methods are run to obtain the optimal parameters $A$ and $B$. Then, using these values, the final excitations $E_{\rm ex,1}(t_{\rm f}, \eta)$ are calculated for a range of perturbation strengths $\eta$ and for the run-times for which the trajectories have been optimised. This yields a grid of excitations depending both on the run-time and the perturbation.  Overlaid over the plots are orange contour lines at a level of $\bar{n} = 0.1$, outlining the areas where we consider excitations to become negligible.  For comparison, \fref{fig:robust}a) shows this plot based on the results of the non-robust full-Hamiltonian optimisation presented earlier in \fref{fig:OneParam17Comp}. The results of the two robustness optimisation methods are shown in \fref{fig:robust}b) and c).

The comparison between the non-robust and robust methods shows that the latter offer a clear improvement, with the exact cost function yielding the most robust result. Even though the non-robust protocols achieve a nearly perfect shortcut at all run-times for unperturbed potentials, they break down for much smaller values of $\eta$ than those optimised for robustness. 

Contrary to the non-robust scheme presented in \sref{sec:numerics}, the two robust methods do not produce negligible excitations at all run-times, as can be seen in the insets of \fref{fig:robust}. These show the final excitations $E_{\rm ex,1}(\tf, \eta=0)$ (cut indicated by a white dashed line), which increase exponentially at shorter run-times, while also showing periodic minima. To quantify the run-time beyond which the robust protocols do not give excitation-free results anymore, the excitations at $\eta = 0$ are fitted with the function $f(t) = \left[a\exp(-bt)\sin^2\left(ct+d\right)\right]$. The non-periodic part $\left[a\exp(-bt)\right]$ can then be interpreted as an envelope function and its intersection time with an energy level of $\bar{n}=0.1$ (indicated by a grey dotted line) is then used to estimate the run-time below which the protocol is not well optimised (indicated by a hatched area). We refer to this time as the critical time $T_{\rm crit}$. For the approximate method, it is found to be $T_{\rm crit, app} = \SI{27.5}{\micro\second}$ (about 12.4 motional cycles), while the exact method performs better at shorter run-times and $T_{\rm crit, ex} = \SI{14.2}{\micro\second}$ ($\sim 6.4$ motional cycles). Note that the definition of $T_{\rm crit}$ is a conservative one, as timings resulting in negligible excitations can be found below $T_{\rm crit}$ due to the periodic nature of the excitations.
Another interesting feature of both robust methods is the existence of stripes with low excitations at constant, periodic run-times, marking configurations where the scheme is ultra-robust even against strong perturbations.

The fact that the robust methods work well despite the simple choice of the cost function is encouraging, as it is plausible that it can be improved in similarly simple ways. One could for example use more free parameters and choose more values of $\eta$ in \eref{eq:cost-app} and \eref{eq:cost-ex}, such that the robustness range is increased.
We conclude that the presented robustness optimisation methods are useful tools to make this STA protocol able to withstand experimental imperfections, at the cost of introducing a limit to the achievable run-times. Having gained sufficient control to perform it in around 6 trap cycles without excess excitations, we apply this result to find timings where the ion motional excitations are swapped.

\section{Cooling characteristics}
\label{sec:cool}

After gaining the ability to optimise the protocol for low residual excitations and optimal robustness over a range of run-times when executed on cold ions, the existence of cooling solutions can now be demonstrated. To this end, we deploy the protocols that were optimised in \sref{sec:robust_opt} and now apply them to the case where one ion is initially hot. In this scenario, motional energy is exchanged between the ions and we therefore expect to find a run-time where a complete energy swap takes place. This timing can be found by calculating the final energy of the initially hot ion after the protocol for a range of run-times. The desired cooling solution is then given by the trajectory at the run-time where this final energy reaches a minimum, indicating an energy swap.

We also explore the effect that varying the inner distance $d_{\rm in}$ has on the timing of the energy swap. As the motional exchange scales strongly with the ion distance, we expect to find cooling solutions at shorter run-times when bringing the ions closer together.
As we have chosen to consider only protocols that leave the ions in separate potential wells, the available quartic confinement in the trap under consideration limits the available range of $d_{\rm in} > d_{\rm c}$. We thus go on to consider how to decrease cooling times by varying $\bmax$.

\subsection{Numerical prerequisites}
\label{ssec:cooling_numerics}

The energy of each ion is calculated in the same way as in \eref{eq:Eex} and \sref{sec:robust_numeric}. However, now we also need to take into account that the ions do not necessarily start in their ground state. The initial conditions are given by setting $p_i(0)$ and $x_i(0)$. This is done by choosing the initial energy $E_{\rm in, i}$ and distributing it onto the kinetic and potential parts of the energy by choosing the initial motional phase $\phi$. A phase $\phi = 0$ is understood to mean that all initial energy is kinetic and thus $x_i(0)$ is equal to the equilibrium position, while $\phi = \pi/2$ implies that initially all energy is stored in the potential and thus $p_i(0) = 0$.
To subsume the effect of $\phi$, we define the average energy $\bar{E}_{\rm ex, i}(\tf, \eta)$, which is obtained by averaging the resulting $E_{\rm ex, i}(\tf, \eta)$ for $\phi \in [0,2\pi]$. For numerical examples in this work, 25 uniformly distributed values of $\phi$ are used.

One ion being initially hot is defined here to mean that it is initialised with a motional energy corresponding to $\nbar = E_{\rm in, i}/\hbar\omega_0 = 10$, which is on the order of the Doppler limit (around 20.5 quanta for $^{40}\mathrm{Ca}^+$ at a mode frequency of $\sim\SI{0.48}{\mega\hertz}$ \cite{Negnevitsky}).

\subsection{Cooling solutions}

We now go on to demonstrate the energy exchange, while varying the inner distance $d_{\rm in}$ from $1.0d_{\rm c}$ to $1.25d_{\rm c}$. Otherwise, the physical constraints are kept identical to those chosen in \sref{sec:numerics} (see table \tref{table:target-values}), namely $\bmax = \SI{0.85e-3}{\newton\per\meter\cubed}$, $d_0 = 5d_{\rm c} = \SI{70.1}{\micro\meter}$ and the mass of a \caf ion. To obtain optimised trajectories, the ansatz \eref{eq:rho_pl} is constructed for each resulting set of physical constraints, leaving the free parameters $A$ and $B$. Optimal values for these, which depend on the run-time $t_{\rm f}$, are then found by applying the exact cost function to initially cold ions as described in the previous section.

\begin{figure}[!ht]
  \centering
    \includegraphics[width=0.8\textwidth]{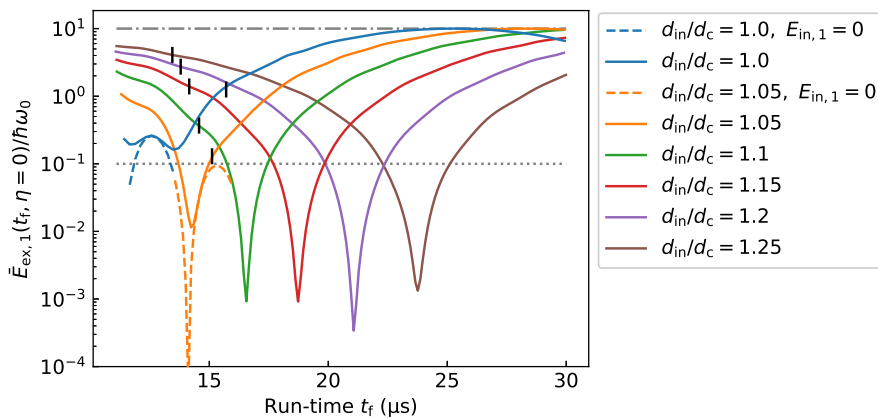}
  \caption{Final energy $\bar{E}_{\rm ex, 1}(\tf, \eta=0)$ (averaged over the motional phase $\phi$) of the ion initially excited by 10 quanta for various choices of $d_\mathrm{in}$. While the physical constraints are otherwise the same as in \tref{table:target-values}, the value of $d_\mathrm{in}$ was varied from $1.0d_{\rm c}$ to $1.25d_{\rm c}$. For each set of physical constraints and at each run-time, the trajectory was first optimised using the exact robust cost function \eref{eq:cost-ex}, before running the scheme on one initially hot ion and one in the ground state. The critical time $T_{\rm crit, ex}$ is also calculated for each set of parameters in the same way as it was for \fref{fig:robust} and the results are indicated by black markers. For the trajectories corresponding to $d_{\rm in}/d_{\rm c} = \{1.0, 1.05\}$, we also plot the final excitations added to an ion initially in the ground state (dashed lines).
  The initial energy is indicated by the dash-dotted line and the target energy of $\nbar = 0.1$ by the dotted line. Note that $\omega_0$ depends on the physical constraints $\{\bmax, d_\mathrm{in}, d_\mathrm{out}, m$\} and thus the normalisation of the final excitations has to be recalculated for each choice of $d_\mathrm{in}$.}
\label{fig:two-cooling}
\end{figure}

Finally, for each run-time, the obtained optimised trajectories are applied to two ions, one of them initialised to $\nbar = 10$ and the other one being in the ground state. The final excitations $\bar{E}_{\rm ex, 1}$ of the initially excited ion are then calculated and the results are shown in \fref{fig:two-cooling}. We observe that the final energy has a minimum far below $\nbar = 0.1$ for each shown value of $d_{\rm in}$ except for $d_{\rm in} = 1.0d_{\rm c}$.  We refer to the run-time of these minima as the cooling time $T_{\rm c}$.
This run-time of complete energy exchange tends to lower values as the ions are brought closer together, as predicted from \eref{eq:exchange}. The final excitations at $T_{\rm c}$ differ from zero due to the schemes causing a finite energy increase even for ground-state ions, as well as due to the simulations only being performed for discrete $\tf$.

By decreasing the minimal distance $d_{\rm in}$, the cooling time is eventually lowered into a range of run-times where the adiabatic shortcut can not be perfectly optimised anymore. The critical time $T_{\rm crit}$ was introduced in \sref{sec:robust_opt} to quantify the onset of this regime. After optimising the trajectories applied in \fref{fig:two-cooling} using the exact method \eref{eq:cost-ex}, $T_{\rm crit,ex}$ is calculated for each of the choices of $d_{\rm in}$ and indicated in \fref{fig:two-cooling} by the black markers. For the lines where $d_{\rm in} \geq 1.1d_{\rm c}$, the cooling time comes to lie above $T_{\rm crit, ex}$. When bringing the ions closer together and thus reducing $T_{\rm c}$ below the critical time, one might expect that the energy swap is not completed to a satisfactory level anymore, due to the adiabatic shortcut potentially adding excitations above 0.1 quanta per ion. This does however not occur in \fref{fig:two-cooling} until $d_{\rm in}=1.0d_{\rm c}$, which is explained by the conservative definition of $T_{\rm crit, ex}$. As one recalls from \fref{fig:robust}, the excitations of two ground-state ions do not increase monotonically with decreasing run-time, but show periodic minima indicating near-perfect shortcuts. However, $T_{\rm crit,ex}$ is defined using the envelope of said excitations, disregarding the minima. It is then indeed the case for $d_{\rm in}=1.05d_{\rm c}$ that the timing of such a trajectory coincides with the timing required for an energy swap, allowing cooling to below 0.1 quanta. Only for $d_{\rm in}=1.0d_{\rm c}$ is this not true anymore. To visualise this effect, we have also plotted the excitations $E_{\rm ex,i}(\tf, \eta=0)$ added to an ion initially in the ground state for these two choices of $d_{\rm in}$ (dashed lines).

\begin{figure}[ht]
    \centering
        \subfloat{\includegraphics[width=0.4\columnwidth]{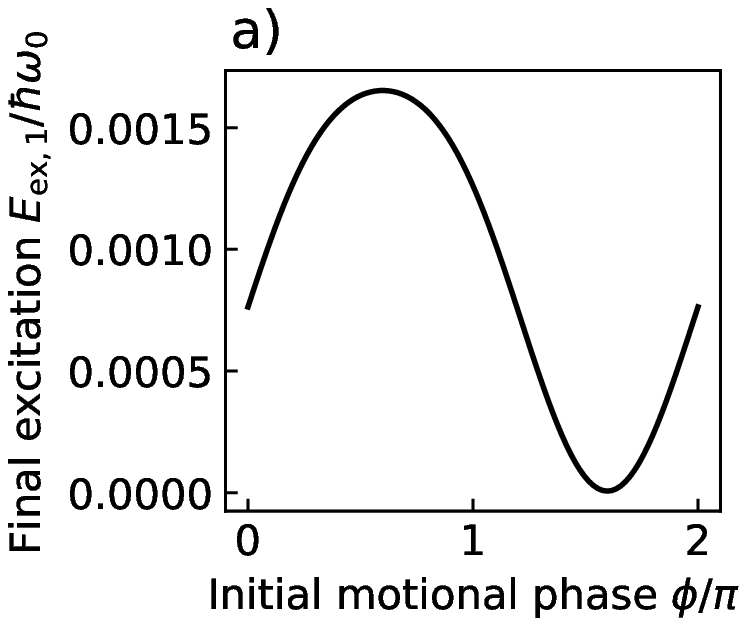}\label{fig:MotionalPhase}}
        \subfloat{\includegraphics[width=0.4\columnwidth]{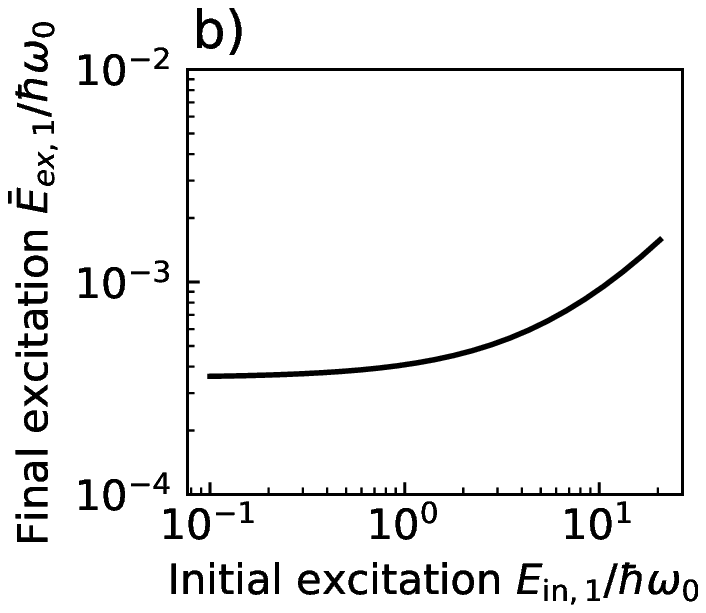}\label{fig:initEnergyDep}}
    \caption{Dependence of the final excitations a) on the initial motional phase and b) the initial excitation. Both plots show the trajectory for $d_\mathrm{in} = 1.1d_{\rm c}$ that provided the cooling solution at $\tf =\SI{16.6}{\micro\second}$, as shown in \fref{fig:two-cooling}. For a), the initially hot ion was initialised with $\nbar = 10$, while this was varied up to the Doppler limit for b).}
\end{figure}

As the final excitation $\bar{E}_{\rm ex, 1}$ is an average over many initial motional phases, we check if this aggregation is justified. We plot the dependence of the non-averaged excitations $E_{\rm ex, 1}$ on $\phi$ in \fref{fig:MotionalPhase}, the example being the protocol with $d_\mathrm{in} = 1.1d_{\rm c}$ at the run-time $\tf = \SI{16.6}{\micro\second}$, which corresponds to the timing of the cooling solution. The dependence has a sinusoidal shape around the average, making it a reasonable replacement.

Furthermore, the final excitation should not depend strongly on choosing an initial energy that differs from the exemplary $\nbar = 10$. This is confirmed in \fref{fig:initEnergyDep}, where $\bar{E}_{\rm ex, 1}$ is shown to be well below $\nbar = 0.1$ for initial excitations ranging up to the Doppler limit.

The existence of trajectories that provide almost complete energy swapping is thus demonstrated, with cooling times $T_{\rm c}$ on the order of 10 motional cycles. While varying $d_{\rm in}$ is a viable option to fine-tune the cooling time, the maximal speed is fundamentally limited by two factors. First, as seen in section \ref{sec:robust_opt}, the integration of robustness into the protocol optimisation introduced the critical time $T_{\rm crit}$, below which no satisfying adiabatic shortcut is found. This limit holds despite the existence of viable trajectories at shorter run-times under certain conditions. Secondly, the maximal quartic confinement $\beta_{\rm max}$ dictates $d_{\rm c}$ and thus the available range of $d_{\rm in}$.

We will therefore go on to analyse the behaviour of the protocol when scaling $\beta_{\rm max}$ in addition to $d_{\rm in}/d_{\rm c}$, expecting that cooling can be achieved faster in a trap that provides a larger quartic confinement.

\subsection{Scaling behaviour with the maximal quartic confinement}

To study the effects of scaling $\beta_\mathrm{max}$ in an easily comparable way, we go over to express all quantities that describe the protocol in a dimensionless form. All energies are expressed on a scale of $\hbar\omega_0$ as before, while values denoting time can be made dimensionless by dividing by a motional cycle $(2\pi)/\omega_0$. The distance trajectory $d(t)$ can be made dimensionless through division by the critical distance $d_{\rm c}$, which is solely dependent on $\bmax$. To set the physical constraints, one then picks $d_0/d_{\rm c}$ ($=5$ in our case) and $d_{\rm in}/d_{\rm c} >1$. Rewritten as such, protocols that share the same $d_0/d_{\rm c}$ and $d_{\rm in}/d_{\rm c}$ can be compared across varying $\bmax$.

To determine how the motional exchange rate $\Omega$ throughout the protocol depends on $\bmax$ and with it the timing of the cooling solutions, we make $\Omega$ dimensionless by division with $\omega_0$ and obtain 
\be
\label{eq:mot_ex_time}
\frac{\Omega}{\omega_0} = \left[\frac{m}{4C_{\rm C}}\omega_0\sqrt{\omega_1\omega_2}d^3\right]^{-1} \ .
\ee
We show in \ref{app:scaling} that this expression scales as $\omega_0\sqrt{\omega_1\omega_2}d^3 \propto \left(\beta_{\rm max}\right)^0$. We then expect that for a given choice of $\{d_0/d_{\rm c}, d_{\rm in}/d_{\rm c}\}$, varying $\beta_{\rm max}$ leaves $\Omega/\omega_0$ invariant and thus the cooling times can always be found at the same number of motional cycles.

To demonstrate this result, we pick four values of $\beta_{\rm max}$ spanning three orders of magnitude. For each of these maximal confinements, the inner distance $d_{\rm in}/d_{\rm c}$ is varied from $1.05$ to $1.25$. The outer distance is chosen to be $d_{\rm out}/d_{\rm c} = 5$ as in all examples so far. For each combination of $\{\bmax, d_{\rm in}/d_{\rm c}\}$, the protocol is optimised for a range of run-times using the exact cost function as described in section \ref{sec:robust_numeric} and the critical time $T_{\rm crit, ex}$ is determined. We then determine the run-time $T_{\rm c}$ leading to a cooling solution in the same way as in the previous subsection. Note that for $\beta_{\rm max} = 1\cdot\beta_{\rm HOA}$, these computations are equal to the ones performed for \fref{fig:two-cooling}.

\begin{figure}[!ht]
  \centering
    \includegraphics[width=1\textwidth]{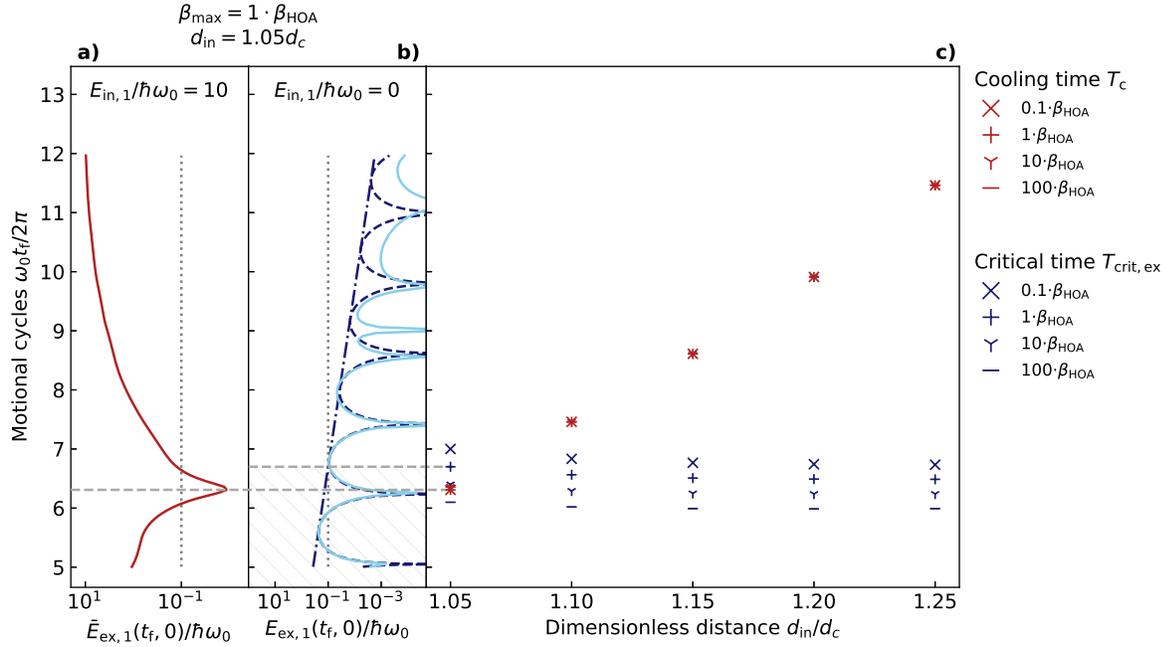}
    \caption{c) Scaling plot showing the cooling solutions $T_{\rm c}$ (red symbols) and the critical time $T_{\rm crit, ex}$ of the exact cost function (blue symbols) depending on the inner distance $d_{\rm in}/d_{\rm c}$ and for four choices of the maximal quartic confinement $\beta_{\rm max} = \{0.1\cdot\beta_{\rm HOA},\ 1\cdot\beta_{\rm HOA},\ 10\cdot\beta_{\rm HOA},\ 100\cdot\beta_{\rm HOA}\}$, where $\beta_{\rm HOA} = \SI{0.85e-3}{\newton\per\meter\cubed}$ is the maximal quartic confinement available in the HOA2 trap, as used in the previous sections. 
    Two insets corresponding to the case of $\beta_{\rm max} = 1\cdot\beta_{\rm HOA}$ and $d_{\rm in}/d_{\rm c} = 1.1$ have been added on the left. a) shows the final excitations of an ion that was initially excited by 10 quanta, exhibiting an energy swap at about 6.3 motional cycles (grey dashed line). b) shows the final excitations of two ions initialised in the ground state after being run through the protocol optimised with the exact cost function. The critical time $T_{\rm min, ex}$ (grey dashed line) is defined as the time when the envelope of the excitations drops below 0.1 quanta.
    Note that the period of a motional cycle $2\pi/\omega_0$ depends on $\{\bmax, d_{\rm in}/d_{\rm c}, d_{\rm out}/d_{\rm c}\}$ through \eref{eq:omega0}. \ref{app:scaling} yields that $\omega_0 \propto \left(\bmax\right)^{\frac{3}{10}}$ when keeping the dimensionless boundary distances constant. For $\bmax = 1\cdot\beta_{\rm HOA}$, $\omega_0$ varies from $2\pi\cdot\left(\SI{0.44}{\mega\hertz}\right)$ to $2\pi\cdot\left(\SI{0.48}{\mega\hertz}\right)$ with the presented choices of $d_{\rm in}/d_{\rm c}$, while going to $\bmax = 100\cdot\beta_{\rm HOA}$ increases the frequency range to $2\pi\cdot\left(\SI{1.76}{\mega\hertz}\right)$ to $2\pi\cdot\left(\SI{1.92}{\mega\hertz}\right)$.}
\label{fig:eq-scaling}
\end{figure}

The cooling times and critical times are plotted in \fref{fig:eq-scaling}c depending on the dimensionless distance. The red symbols show the cooling times $T_{\rm c}$ obtained for all the combinations of $\{\bmax, d_{\rm in}/d_{\rm c}\}$. No dependence on $\bmax$ is observable as expected. Choosing a smaller inner distance leads to a faster overall exchange, as already discussed in \fref{fig:two-cooling}. The critical times $T_{\rm crit, ex}$ are depicted by the blue symbols and only a weak dependence on both the quartic confinement and the inner distance is observed, allowing us to state that the exact cost function produces well-optimised trajectories down to $6-7$ motional cycles.

As already observed in \fref{fig:two-cooling}, cooling solutions are also found which have shorter duration than the critical time of the exact cost function $T_{\rm crit, ex}$. This is shown in detail again in \fref{fig:eq-scaling}a) and \ref{fig:eq-scaling}b) through the example of the trajectories corresponding to $\bmax = 1\cdot\beta_{\rm HOA}$ and $d_{\rm in}/d_{\rm c} = 1.05$. \Fref{fig:eq-scaling}a) depicts the final excitations of an ion that was initially excited by 10 quanta, depending on the run-time. \Fref{fig:eq-scaling}b) on the other hand shows the excitations caused by the protocol when both ions are initially in their ground state, together with a fit to the simulated data and the excitation envelope as described in \sref{sec:robust_opt}. 
It becomes apparent that this particular cooling solution is enabled by the fact that the excitations in \fref{fig:eq-scaling}b) show a minimum at 6.3 motional cycles, despite the envelope already indicating far higher excitations.

The results presented in \fref{fig:eq-scaling} provide an easy way to select the parameters of a desired cooling protocol. The energy of two ions can be swapped at best within $6-7$ motional cycles. The ideal choice of the inner distance is then also immediately clear as being $\sim 1.05d_{\rm c}$, providing a cooling solution at $\sim 6.3$ motional cycles. As calculated in \ref{app:scaling}, the trap period $2\pi/\omega_0$ scales with $\bmax^{-3/10}$ and the same is then true for the cooling solutions $T_{\rm c}$ in absolute time. However, also consider that we have chosen $d_{\rm 0}/d_{\rm c} = 5$ in all presented examples. If a larger value is selected in a given implementation, we would expect the cooling minima and the critical times to tend to longer run-times, as the ions spend more time far away from each other and are more strongly accelerated.

\section{Suppression of motional exchange by an additional homogeneous electrical field}
\label{sec:resViol}

As noted in the introduction, the motional exchange is a resonant effect with respect to the ion frequencies. Since an additional homogeneous electrical field introduces a mismatch of the potential curvatures of each ion as calculated in subsection \ref{ssec:mismatch}, this resonance condition is not perfectly observed anymore. We therefore proceed to estimate the range of tolerable perturbations $\eta$ such that cooling is still achieved.

Due to being a resonance effect, we choose an ansatz in which the final excitation $E_{\rm ex,1}$ of the initially hot ion has a Lorentzian shape with respect to the curvature mismatch $\delta\omega$:
\be
\label{eq:lorentz}
E_{\rm ex,1} = E_{\rm in,1}\left(1 - \frac{1}{1 + \left(k\frac{\delta\omega}{\Omega}\right)^2}\right),
\ee
where $E_{\rm in,1}$ is the initial ion energy, $\Omega$ is the exchange frequency \eref{eq:exchange} and $\delta\omega$ the frequency mismatch \eref{eq:freq_mis}, while the factor $k$ is a constant that is to be determined. 
From this, the term $\left(k\frac{\delta\omega}{\Omega}\right)$ is calculated as 
\be
\label{eq:width}
k\frac{\delta\omega}{\Omega} = \frac{12k\beta d^4d_{\rm in}}{C_{\rm C}}\eta \sim \overbrace{\frac{12k\bmax d_{\rm in}^5}{C_{\rm C}}}^{\left(\eta_{1/2}\right)^{-1}}\eta 
\ee
where we have replaced the time-dependent values $d$ and $\beta$ by their values at $t_{\rm f}/2$, under the assumption that most of the energy exchange happens at that point in time. We understand $\eta_{1/2}$ to be the Lorentzian half-width (HWHM) with respect to the perturbation $\eta$. Using \eref{eq:critDist}, we obtain the simple expression

\be
\label{eq:etaHWHM}
\eta_{1/2} = \frac{1}{24k(d_{\rm in}/d_{\rm c})^5}
\ee
which depends on $k$ and on how close one brings the ions to the critical point, but not on $\bmax$ itself. This means that the required experimental accuracy in terms of the perturbation parameter $\eta$ can not be reduced by using a trap with larger quartic confinement. However, it can be maximised by choosing a minimal value for $d_{\rm in}/d_{\rm c}$. Thus the goal of swapping the motional energy as fast as possible is compatible with finding a cooling solution that is maximally insensitive to additional homogeneous fields.

\begin{figure}[!htbp]
  \centering
    \subfloat{\includegraphics[width=0.5\columnwidth]{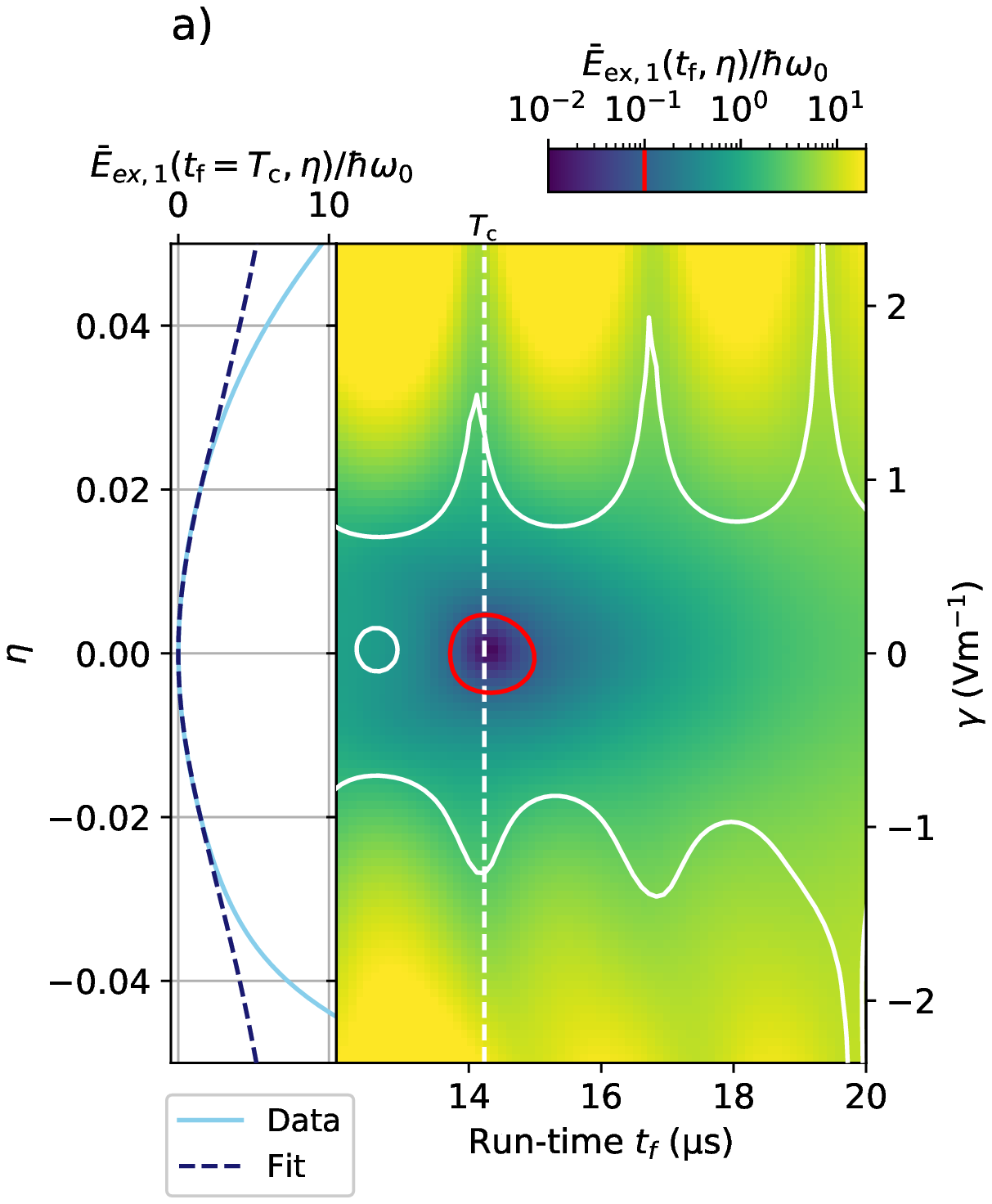}\label{fig:resonance_105}}
    \subfloat{\includegraphics[width=0.5\columnwidth]{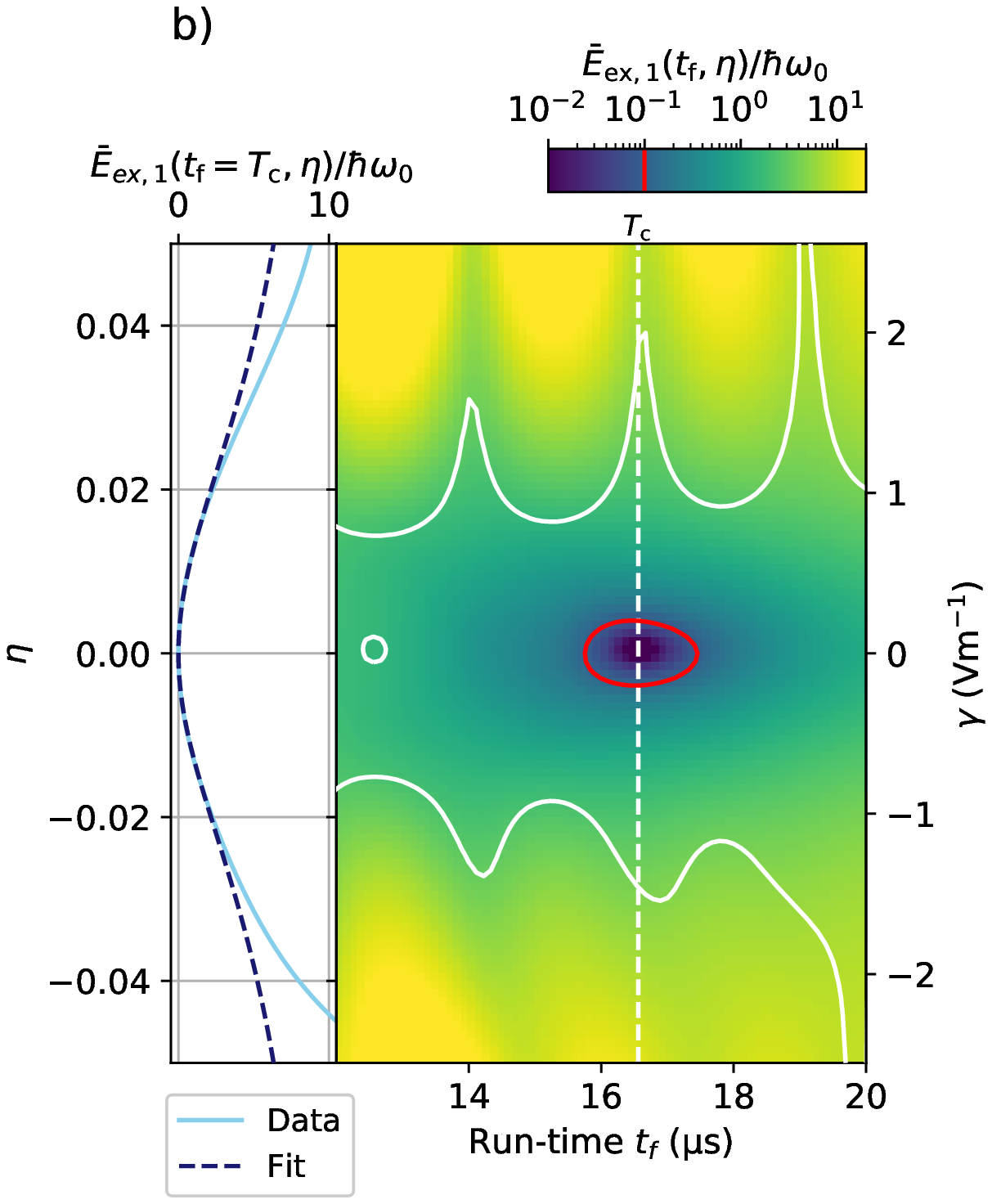}\label{fig:resonance_110}}
  \caption{Final excitation $\bar{E}_{\rm ex,1}$ of the initially hot ion depending on the run-time $t_{\rm f}$ and the perturbation $\eta$ for an inner distance of a) $1.05d_{\rm c}$ and b) $1.10d_{\rm c}$. The other physical constraints of the trajectories are $\bmax = \SI{0.85e-3}{\newton\per\meter\cubed}$ and $d_{\rm 0}/d_{\rm c} = 5$ and the free parameters $A$ and $B$ are reused from the optimisations performed for \fref{fig:two-cooling}, meaning that they were optimised using the exact cost function. The red contour line marks the area where $\bar{E}_{\rm ex,1}$ decreases below 0.1 quanta. Overlaid over the plots are further contour lines similar to those of \fref{fig:robust}, marking the area where the scheme excites an ion by less than 0.1 quanta, if applied to ground-state ions. The insets to the left of the plots show a cut through the data (indicated by the dashed white line) at the cooling solution a) $t_{\rm f} = T_{\rm c} = \SI{14.2}{\micro\second}$ and b) $t_{\rm f} = T_{\rm c} = \SI{16.6}{\micro\second}$, together with a fit of the Lorentzian \eref{eq:lorentz} to part of the data that fulfils $\bar{E}_{\rm ex,1}/\hbar\omega_0 < 2$, in which values of a) $k=0.678$ and b) $k=0.679$ are obtained. The homogeneous electrical field values $\gamma$ that corresponds to a given $\eta$ is shown on separate scales on the right.}
\label{fig:eq-resonance}
\end{figure}

To illustrate these results, \fref{fig:eq-resonance} recreates the robustness plots from \sref{sec:robust_opt}, but with one ion initially excited by 10 quanta. This is done for the two inner distances yielding the fastest cooling, $d_{\rm in}/d_{\rm c} = 1.05$ and $d_{\rm in}/d_{\rm c} = 1.10$. For each run-time shown, we apply the trajectory that was optimised using the exact cost function for \fref{fig:two-cooling}, but with a range of perturbations parametrised by $\eta$. For each $\eta$, one ion is initialised with an excitation of 10 quanta and the equations of motion are calculated while keeping the perturbation constant throughout the scheme. The resulting final excitation $\bar{E}_{\rm ex,1}(\tf, \eta)$ is then calculated in the same way as described in \sref{ssec:cooling_numerics} and plotted on a grid of run-times and perturbations. Note that $\bar{E}_{\rm ex,1}$ is again an average over the initial motional phase. The red contour line shows the area where the final excitation of the initially hot ion decreases below $\bar{n} = 0.1$.
For comparison, we overlay another contour line in white, marking the area where the trajectories, applied to ground-state ions, cause a final excitation of the ion below 0.1 quanta.
To show the resonance effect, we take a cut though the data at the cooling time $T_{\rm c} = \SI{14.2}{\micro\second}$ and $T_{\rm c} = \SI{16.6}{\micro\second}$, respectively, which are familiar from \fref{fig:two-cooling}. The cuts are indicated by the white dashed lines. This yields the insets on the left, showing the excitation $\bar{E}_{\rm ex,1}$ depending on the perturbation.

To the data of these cuts, we fit the Lorentzian \eref{eq:lorentz}. Specifically, we insert \eref{eq:width} into the Lorentzian and find the optimal value for the remaining parameter $k$ by applying the Python method \textit{curve\_fit} on the central part of the data that fulfils $\bar{E}_{\rm ex,1}/\hbar\omega_0 < 2$. The resulting fit functions are also displayed in the left insets in \fref{fig:eq-resonance}, showing a good match to the data in the central peak. A value of $k=0.679$ was obtained in a), corresponding to $\eta_{1/2} = 0.048$ and $k=0.678$ in b), corresponding to $\eta_{1/2} = 0.038$. In order to assure cooling to better than $\bar{n}=0.1$, we therefore find that $\eta$ must not exceed values of $\pm0.0048$ and $\pm0.0038$ respectively. The scheme that brings the ions closer together thus yields better tolerance to stray fields in addition to faster cooling, confirming \eref{eq:etaHWHM}. We repeat this analysis for $d_{\rm in}/d_{\rm c} = \{1.15,1.2,1.25\}$, obtaining values of $k$ close to $2/3$ as well. Thus we can regard the Lorentzian ansatz and the form of $\eta_{1/2}$ in \eref{eq:etaHWHM} as a reasonable estimate for the resonance behaviour close to $\eta=0$, despite the crude approximation that led to this result.

The nature of the motional exchange imposes much stricter conditions on the maximally tolerable perturbation $\eta$ than the in-and-out transport of two ground-state ions, as can be seen from the contour lines in \fref{fig:eq-resonance}. To achieve the desired motional exchange, $\eta$ can not exceed values of $\pm 0.0048$, corresponding to an electric field of $\pm\SI{0.22}{\volt\per\meter}$. Previously, a stray field calibration in steps of $\SI{0.1}{\volt\per\meter}$ was reported \cite{Ruster2014}, although in a trap where the available $\bmax$ was about 17 times lower than in the one considered here.

Contrary to $\eta_{1/2}$, the corresponding homogeneous field strength $\gamma_{1/2}$ does scale with $\bmax$. From the conversion given in \eref{eq:lambdaParameter} and the results of \ref{app:scaling}, we obtain that
\be
\gamma_{1/2} \propto (\bmax)^{2/5} \ .
\ee
The tolerable field strength thus increases with the quartic confinement, whereas $\eta_{1/2}$ is invariant. This allows us to compare this result to experimental demands in other experiments. $\bmax$ scales with the overall trap size scale $a$ as $\bmax \propto a^{-4}$ \cite{Home2006}.

The tolerable value of $\eta$ also strongly depends on the desired initial and final excitation level. This can be seen by rewriting \eref{eq:lorentz} and \eref{eq:width} to 
\be
\eta = \eta_{1/2}\sqrt{\frac{E_{\rm ex,1}/E_{\rm in,1}}{1-E_{\rm ex,1}/E_{\rm in,1}}} \ .
\ee
For example, if one intends to cool from 1 to 0.1 quanta instead of from 10 to 0.1 quanta as shown in \fref{fig:eq-resonance}, the tolerable range of $\eta$ increases by a factor of 3.3.

Note finally that the available range of parameters also depends on the choice of the optimisation cost function \eref{eq:cost-ex}. In a previous iteration of this work, we had optimised the schemes with ground-state ions using $\eta = 0.03$ in the cost functions \eref{eq:cost-app} and \eref{eq:cost-ex}, causing $T_{\rm crit,ex}$ to lie at 10 motional periods. In that configuration, inner distances larger than $1.05d_{\rm c}$ had to be chosen, resulting in tolerable perturbations of $|\eta| < 0.0020$, making worse use of the available range of robustness.

\section{Cooling ions of unequal mass}
\label{sec:unequal}

Thus far, we have only considered exchange protocols where both ions are of equal mass. However a number of recent works have used ion chains containing ions of different mass \cite{Negnevitsky2018,Ballance2015,Tan2015, Meiners2018,Brewer2019,Hannig2019}, justifying the need to implement cooling in such configurations. We therefore aim to generalise the cooling protocol to ions of unequal masses $m_1 \neq m_2$ in this section.

When trying to extend the cooling scheme to ions of unequal masses $m_1 \neq m_2$, a fundamental problem arises: if we keep using a symmetric potential such as the harmonic-quartic double-well potential \eref{eq:quartPot}, then the ions will always sit at symmetric equilibrium positions. The frequencies resulting from the potential curvatures given by \eref{eq:co-motFreq} are then not equal, as the second derivative of the potential at the ion position is the same for both ions, but the mass is not. The resonance condition $\omega_1 = \omega_2$ for motional exchange to take place is thus violated. As noted in \cite{Lizuain2017}, we are furthermore unable to find decoupled dynamical normal modes for the unequal mass case when using a harmonic-quartic potential.

This is mitigated by introducing an asymmetric term to the potential $V_{\rm el}$, which can be used to force the motional frequencies to be equal throughout the scheme. The simplest choice is to add a linear term leading to the trapping potential

\begin{equation}
\label{eq:potUnequal}
V_\mathrm{el}(x,t) = \gamma(t)x + \alpha (t)x^2 + \beta (t)x^4.
\end{equation}
The linear term is reminiscent of section \ref{sec:robust}, where such a linear term appeared as an undesirable perturbation. Here in contrast, the additional field is intentional. The dynamical normal modes are derived in the same way as in the equal mass case and a detailed account is given in \ref{app:unequal}. The full Hamiltonian is again separated into two harmonic oscillator Hamiltonians and we find that the auxiliary functions $\rho_\pm$ need to fulfil \eref{eq:equalAuxODE:rho} and the commutation is ensured by the same boundary conditions \eref{eq:BC} as before. If the COM-mode frequency is chosen to be constant again, this means that the ansatz for $\rho_+$ given in \eref{eq:rho_pl} can be reused, as well as $\rho_- = 1$. The protocol is then defined by $\rho_\pm$ and \eref{eq:equalAuxODE:rho} as

\be
\label{eq:unequalProtocol}
d(t) &= \sqrt[3]{\frac{4C_{\rm C}}{\sqrt{m_1m_2}(\Omega_+^2 - \Omega_-^2)}}\\
\beta(t) &= \frac{m_1+m_2}{8d^2}\left(\Omega_+^2 + \Omega_-^2\right) - \frac{2C_{\rm C}}{d^5}\\
s(t) &= \frac{m_2-m_1}{48\beta d}\left(\Omega_+^2 + \Omega_-^2\right)\\
\alpha(t) &= \frac{C_{\rm C}}{d^3}- \frac{\beta d^2}{2} - 6\beta s^2\\
\gamma(t) &= -2\alpha s - 2\beta \left(\frac{3}{2}d^2s + 2s^3\right) \ .
\ee

\begin{figure}[!htb]
    \centering
        \subfloat{\includegraphics[width=0.385\columnwidth]{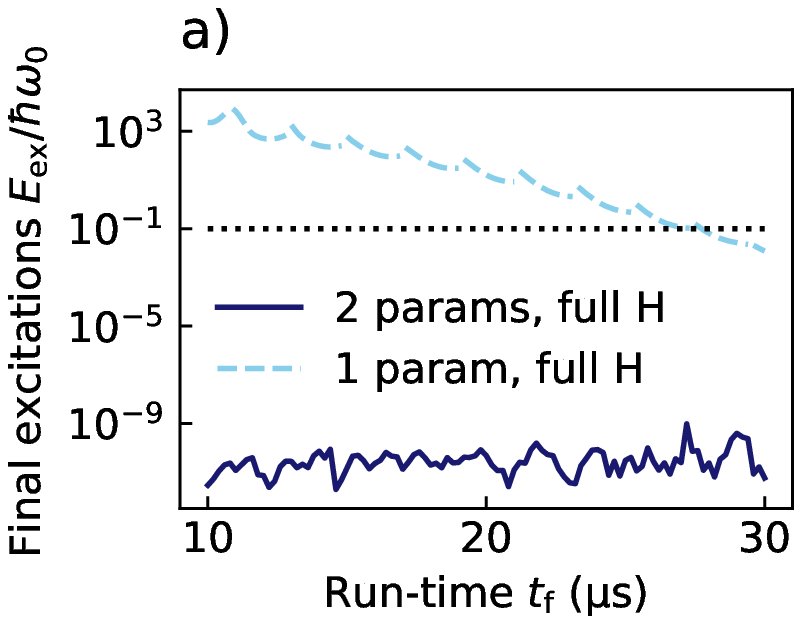}\label{fig:uneq-optim}}
        \subfloat{\includegraphics[width=0.615\columnwidth]{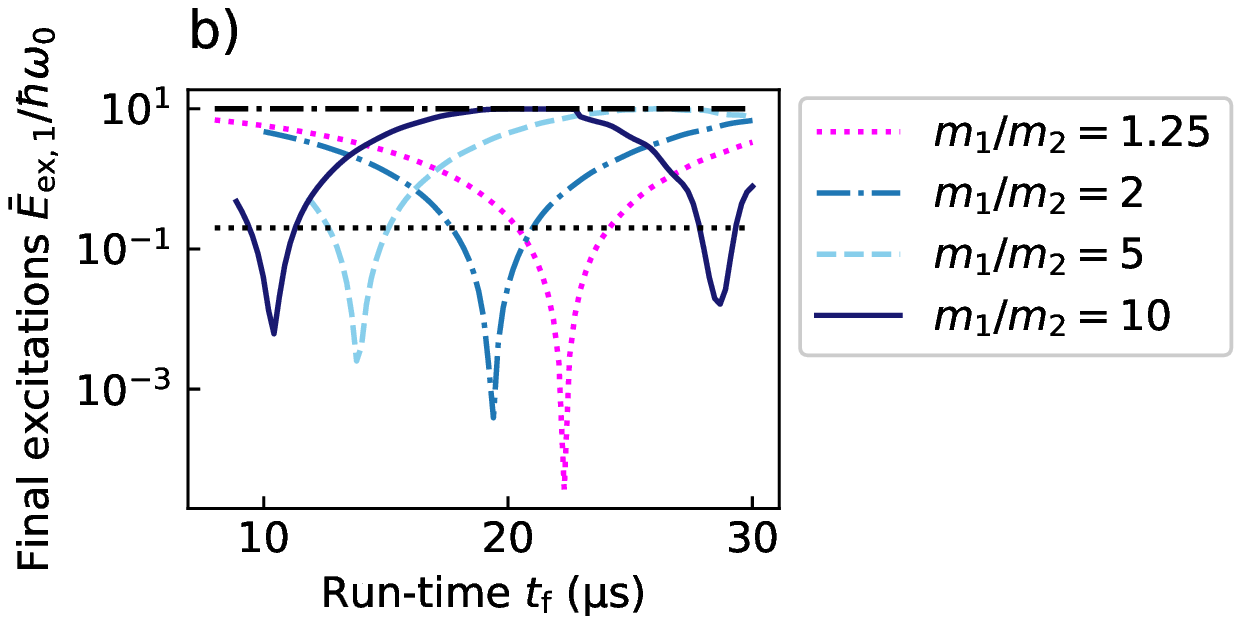}\label{fig:uneq-results-cooling}}
    \caption{Optimisation of the unequal-mass protocols and cooling solutions for a range of mass rations. (a) shows the total excitations $E_{\rm ex,1}$ after optimising the unequal-mass protocol. For each run-time, the optimal free parameters were found by minimising the final energy of the full Hamiltonian. This was done once with just one free parameter ($A$, while keeping $B=0$) and once with both. The excitations are normalised by the energy of a single quanta $\hbar\omega_0$, where $\omega_0 = 2\pi\cdot\left(\SI{0.55}{\mega\hertz}\right)$ for these physical constraints. (b) shows the final excitations $\bar{E}_{\rm ex, 1}$ of the first ion, which has the mass of \caf and is initially excited by 10 quanta. This is depicted for several mass ratios $m_1/m_2 = \{1.25, 2, 5, 10\}$, where the free parameters $A$ and $B$ were optimised for each choice of mass. The excitations are normalised by the energy of a single quanta $\hbar\omega_0$, where $\omega_0$ changes with the masses and ranges from $2\pi\cdot\left(\SI{0.51}{\mega\hertz}\right)$ to $2\pi\cdot\left(\SI{0.62}{\mega\hertz}\right)$.}
    \label{fig:uneq-results} 
\end{figure}

In the same way as in the equal mass case before, we now optimise the total final excitations $E_{\rm ex}$ calculated with the full Hamiltonian by varying the free parameters in $\rho_+$. \Fref{fig:uneq-optim} shows the final excitations of two ground-state ions after optimising only $A$ vs. optimising both $A$ and $B$ for each run-time. The physical constraints are chosen to be as in table \ref{table:target-values}, but with the second ion (coolant) being only half the mass of the first, which is chosen to be that of $\caf$. 

When only using one parameter, the excitations increase exponentially with decreasing run-time despite using the exact Hamiltonian, while negligible excitation levels are reached when optimising both $A$ and $B$. This can be compared to the results for non-robust protocols with ions of equal mass in \sref{sec:sec2}, where only one free parameter had to be optimised to achieve comparably optimal trajectories. Since $q_-=0$ does not hold anymore in this version of the scheme, the parameter search needs to satisfy the boundary conditions for both $q_+$ and $q_-$ and two free parameters are required.

The achievable motional exchange in the unequal-mass case is demonstrated in \fref{fig:uneq-results}b). The first ion is again chosen to have the mass of $\caf$, while the mass $m_2$ of the second is lowered. The two-parameter optimisation is run for a wide range of mass ratios of $m_1/m_2 = \{1.25, 2, 5, 10\}$. Shown is the final energy $\bar{E}_{\rm ex,1}$ of the first ion, initially excited by 10 motional quanta, exhibiting minima where the two ions exchange their motional energy almost completely. The cooling solutions are shifted to shorter run-times with increasing mass ratio. This is explained by the mass dependence of the exchange frequency $\Omega$.

We have therefore demonstrated the existence of trajectories that provide almost complete energy swapping for two ions with mass ratios up to 10. Note that while no robustness towards experimental imperfections has been built into the scheme in this work, this could be generalised in a straightforward way from 
\sref{sec:robust}.

\section{Conclusion and outlook}
\label{sec:conclusion}

Our work demonstrates the possibility to perform fast cooling of a trapped ion by transient interaction with a pre-cooled ancillary ion. For $\caf$ ions, we obtain trajectories which achieve this in 6.3 trap periods, corresponding to \SI{14.2}{\micro\second} for a trap frequency of \SI{0.44}{\mega\hertz} in a currently operated ion trap. The required electric field control is at levels similar to those reported in recent experiments. Similar methods to those reported here would be applicable to cooling of radial modes of motion, although it appears challenging to simultaneously cool both axial and radial modes.

In addition to direct application to cooling, dynamic resonant swapping of motional states would also be a key ingredient for quantum information using oscillator codes rather than internal state qubits \cite{Fluhmann2019}. In addition to the requirements above, this would require consideration of the phase control of the oscillator through the transport.

\ack
This project has received funding from the European Research Council (ERC) under the European Union’s Horizon 2020 research and innovation program grant agreement No 818195, as well as from Intelligence Advanced Research Projects Activity (IARPA), via the US Army Research Office grant W911NF-16-1-0070.

\appendix
\section{Lewis-Riesenfeld invariants for harmonic oscillators}
\label{app:LRtheory}

In this Appendix we state the Lewis-Riesenfeld invariant for a general harmonic oscillator and its eigenstates. This will also yield the energy expectation values of the normal modes as well as the commutation conditions stated in \sref{sec:sta}. We follow the treatment given in \cite{Torrontegui2011}.

A one-dimensional harmonic oscillator Hamiltonian has the form
\begin{equation}
\label{eq:knownHam}
H=\frac{p^2}{2M} - F(t)x + \frac{M}{2}\Omega^2(t)x^2 \ ,
\end{equation}
where $\{p,x\}$ are the canonical coordinates, $M$ is the particle mass, $\Omega(t)$ the time-dependent oscillation frequency and $F(t)$ a force term.

By completing the square and adding a purely time-dependent and thus physically irrelevant term, \eref{eq:knownHam} becomes

\be
\label{eq:trappedHam}
H_{\mathrm{HO}}=\frac{p^2}{2M} + \frac{M}{2}\Omega^2(t)\left(x - \frac{F(t)}{M\Omega^2(t)}\right)^2,
\ee
and we see that this Hamiltonian has the same form as the dynamical normal mode Hamiltonians $H_\mathrm{HO}^{(\pm)}$ obtained in \eref{eq:equalHam}, once with $F(t) = -\sqrt{m/2}\ddot{d}$ and once with $F(t) = 0$ and the mass $M$ set to 1.

The corresponding invariant is defined through two auxiliary functions $\rho(t)$ and $q(t)$ and is given by 

\be
\label{eq:generalInvariant}
I = \frac{1}{2M}\left[\rho(p-M\dot{q}) - M\dot{\rho}(x-q)\right]^2 + \frac{1}{2}M\Omega_0^2\left(\frac{x-q}{\rho}\right)^2
\ee
with $\Omega_0 = \Omega(t=0)$ being the initial frequency.
The auxiliary functions $\rho(t)$ and $q(t)$ need to fulfil the ordinary differential equations (ODE)

\be
\label{eq:auxODEGeneral}
\ddot{\rho}+\Omega^2\rho &= \frac{\Omega_0^2}{\rho^3} \\
\ddot{q} + \Omega^2q &= \frac{F}{M},
\ee
but can be freely chosen otherwise. Again, this is the general form of the ODE given in \eref{eq:equalAuxODE}. Note that we suppress the time-dependence of $\Omega(t)$, $\rho(t)$, $q(t)$ and $F(t)$.

The eigenvectors $\ket{n;t}$ of the invariant, which are needed for Lewis-Riesenfeld theory, are known and their position-space wave functions $\Phi_n(x,t)$ are given by 

\be
\label{eq:invEigenvectors}
\Phi_n(x,t) = \braket{x}{n;t} = \frac{1}{\rho^{1/2}}e^{\frac{iM}{\hbar}\left[\dot{\rho}x^2/2\rho + (\dot{q}\rho - q\dot{\rho})x/\rho\right]}\mathfrak{H}_n\Big(\underbrace{\frac{x-q}{\rho}}_{\equiv\sigma}\Big)
\ee
where the $\mathfrak{H}_n(\sigma)$ are solutions of the instantaneous initial Schrödinger equation with quantum number $n$

\be
\label{eq:instSchrodinger}
\left[-\frac{\hbar^2}{2M}\frac{\partial^2}{\partial \sigma^2}+ \frac{1}{2}M\Omega_0^2\sigma^2\right]\mathfrak{H}_n(\sigma) = \hbar(n+1/2)\Omega_0\mathfrak{H}_n(\sigma).
\ee
This is simply the Schrödinger equation for a static HO in the normalised coordinate $\sigma$, for which the solutions are given by the Hermite functions 

\be
\label{eq:instHarmSolutions}
\mathfrak{H}_n(\sigma) = \frac{1}{\sqrt{2^n n!}}\left(\frac{M\Omega_0}{\pi\hbar}\right)^{1/4}e^{-\frac{M\Omega_0\sigma^2}{2\hbar}}H_n\left(\sqrt{\frac{M\Omega_0}{\hbar}}\sigma\right)
\ee
where $H_n$ are the Hermite polynomials.
The condition for invariant and Hamiltonian to commute at a given point in time $t$ can already be derived from this, as commutation means that the eigenstates of $I(t)$ and the instantaneous Schrödinger solutions of $H_{\rm HO}(t)$ coincide. By requiring $\Phi_n(x,t) = \mathfrak{H}_n(x)$, \eref{eq:invEigenvectors} yields the conditions $q(t) = \dot{q}(t) = \dot{\rho}(t) = 0$ and $\rho(t) = 1$.

Now that the eigenstates are known explicitly, the instantaneous energies can be calculated to be

\be
\label{eq:genInstEnergies} 
E_{n}(t) = \bra{n;t} H_\mathrm{HO} \ket{n;t} = &\frac{\hbar(2n+1)}{4\Omega_{0}}\left(\dot{\rho}^2 + \Omega(t)^2\rho^2+\frac{\Omega_{0}^2}{\rho^2}\right)\\
&+ \frac{M}{2}\dot{q}^2 + \frac{M}{2}\Omega(t)^2\left(q - \frac{F(t)}{M\Omega^2(t)}\right)^2\nonumber \ .
\ee
Note that the phases $\alpha_n(t)$ from \eref{eq:superpos} are irrelevant in obtaining the energies.

\section{Scaling the maximal quartic confinement}
\label{app:scaling}

We want to determine the behaviour of the ion distance $d(t)$ and the potential curvatures $\omega_i(t)$ when scaling the value of the maximal quartic confinement $\beta_{\rm max}$. For this we rewrite the distance as 

\begin{equation}
\label{eq:d-beta-dep}
d(t) = \underbrace{\frac{d(t)}{d_{\rm c}}}_{\equiv D(t)}d_{\rm c} = D(t)\left(\frac{2C_{\rm C}}{\beta_{\rm max}}\right)^{\frac{1}{5}}\; ,
\end{equation}
where $D(t)$ is the dimensionless distance trajectory that we assume to be the same for all values of $\beta_{\rm max}$. This turns out to be true after comparing the results of the numerical optimisations. In the examples shown in \fref{fig:eq-scaling}, $D(t)$ varied from $D(0)=5$ to $D(t_{\rm f}/2)=\{1.0, 1.15, 1.20, 1.25\}$ and back to $D(t_{\rm f})=5$. We thus conclude that the distance scales as $d(t) \propto \left(\beta_{\rm max}\right)^{-1/5}$.

The quartic potential can be easily rewritten in the same way to 
\be
\label{eq:beta-beta-dep}
\beta(t) = \underbrace{\frac{\beta(t)}{\beta_{\rm max}}}_{\equiv B(t)}\beta_{\rm max} = B(t)\beta_{\rm max}\; ,
\ee
where we defined the dimensionless quartic potential term $B(t)$.

Using these results, the scaling of the potential curvatures $\omega_i$ can be found by rewriting \eref{eq:co-motFreq} to 

\be
m\omega_i^2(t) &=& 2\alpha + 3\beta d^2 + \frac{2C_C}{d^3} = 2\beta d^2 + \frac{4C_C}{d^3}\nonumber\\
&=& 2B(t)D^2(t)\left(2C_{\rm C}\right)^{\frac{2}{5}}\left(\beta_{\rm max}\right)^{\frac{3}{5}} + 2\left(2C_{\rm C}\right)^{\frac{2}{5}}\left(\beta_{\rm max}\right)^{\frac{3}{5}} \; ,
\ee
where we have used \eref{eq:distance} for the second equality. The curvatures therefore scale as $\omega_i \propto \left(\beta_{\rm max}\right)^{3/10}$. One can easily see from \eref{eq:nm_freq:-} and \eref{eq:nm_freq:+} that the same is true for $\Omega_-$ and $\Omega_+$.

The scaling of the motional exchange time in \eref{eq:mot_ex_time} can now easily be proven by collecting the scaling of the constituting terms. We find that

\be
\frac{m}{4C_{\rm c}}\omega_0\sqrt{\omega_1\omega_2}d^3 &\propto \omega_i(0)\omega_i(t) d^3(t) \propto \left(\beta_{\rm max}\right)^{\frac{3}{10}}\left(\beta_{\rm max}\right)^{\frac{3}{10}}\left(\beta_{\rm max}\right)^{-\frac{3}{5}}\nonumber\\ &\propto \left(\beta_{\rm max}\right)^{0}\; .
\ee

The homogeneous field $\gamma$ and the perturbation parameter $\eta$ are related by \eref{eq:lambdaParameter} and scale as

\be
\gamma \propto \Omega_-^2d_{\rm in} \propto (\bmax)^{\frac{2\cdot3}{10}}(\bmax)^{\frac{-1}{5}} = (\bmax)^\frac{2}{5} \ .
\ee

\section{Dynamical normal modes for ions of unequal mass}
\label{app:unequal}

The equilibrium positions fulfil 

\be
\label{eq:unequalEqConds}
\frac{\partial V}{\partial x_i}\Bigg{|}_{x_i^{(0)}} = \gamma+ 2\alpha x_i^{(0)} + 4\beta\left(x_i^{(0)}\right)^3 - \frac{(-1)^iC_C}{x_2^{(0)}-x_1^{(0)}}=0 \ ,
\ee
where $i=\{1,2\}$. As in \sref{sec:robust_opt}, we introduce the shifted parametrisation $x_1^{(0)} = s-\frac{d}{2}$ and $x_2^{(0)} = s + \frac{d}{2}$.

The mass-weighted Hessian $K$ that needs to be diagonalised is given by
\be
K = 
 \left(\begin{array}{cc}
  \frac{2\alpha + 12\beta \left(x_1^{(0)}\right)^2 + \frac{2C_C}{d^3}}{m_1} & -\frac{2C_C}{\sqrt{m_1m_2}d^3} \\
  -\frac{2C_C}{\sqrt{m_1m_2}d^3} & \frac{2\alpha + 12\beta \left(x_2^{(0)}\right)^2 + \frac{2C_C}{d^3}}{m_2}
 \end{array}\right) \ .
\ee
Note then that having equal potential curvatures at all times is an equivalent condition to having equal diagonal entries $K_{11} \stackrel{!}{=} K_{22}$. Enforcing this condition, the matrix $K$ takes a symmetric form and is thus easily diagonalised with same constant eigenvectors as in \sref{sec:sta}
 
\be
v_\pm &= \frac{1}{\sqrt{2}}\left(\begin{array}{c}
  1 \\
  \mp1
 \end{array}\right) \ .
\ee

The eigenvalues are
\be
\label{eq:unequalEigenvals}
\Omega_\pm^2 &= \frac{1}{m_1}\left[2\alpha + 12\beta \left(x_1^{(0)}\right)^2 + \frac{2C_C}{d^3}\left[1 \pm \sqrt{\frac{m_1}{m_2}}\right]\right]\nonumber\\
&= \frac{1}{m_2}\left[2\alpha + 12\beta \left(x_2^{(0)}\right)^2 + \frac{2C_C}{d^3}\left[1 \pm \sqrt{\frac{m_2}{m_1}}\right]\right] \ .
\ee
The change of variables to normal-mode coordinates is given by

\be
\label{eq:unequalCoordChangeMat}
 A(t) = \frac{1}{\sqrt{2}}\left(\begin{array}{cc}
  \sqrt{m_1} & \sqrt{m_2}\\
  -\sqrt{m_1} & \sqrt{m_2}
  \end{array}\right) \ ,
 \ee
making the position coordinates 
\be
\label{eq:unequalNMPosCoords}
\left(\begin{array}{c}
  X_- \\
  X_+
 \end{array}\right) &=&  A(t)\left(\begin{array}{c}
  x_1 - (s-\frac{d}{2})\\
  x_2 - (s+\frac{d}{2})
 \end{array}\right)
\ee
and the momentum coordinates
\be
\label{eq:unequalNMMomCoords}
\left(\begin{array}{c}
  P_- \\
  P_+
 \end{array}\right) &=&  \left(A^T(t)\right)^{-1}\left(\begin{array}{c}
  p_1\\
  p_2
 \end{array}\right) - A(t)\left(\begin{array}{c}
  \dot{s}-\frac{\dot{d}}{2}\\
  \dot{s}+\frac{\dot{d}}{2}
 \end{array}\right) \ .
\ee
This finally gives us the dynamical normal mode Hamiltonian in the explicit form

\be
\label{eq:unequalHam}
\fl H_\mathrm{2HO}=& H_\mathrm{HO}^{(+)} + H_\mathrm{HO}^{(-)}\\
\fl H_\mathrm{HO}^{(\pm)} =& \frac{{P}_\pm^2}{2} + \frac{1}{2}\Omega_\pm^2\left({X}_\pm + \frac{\sqrt{m_2} \mp \sqrt{m_1}}{\sqrt{2}}\frac{\ddot{s}}{\Omega_\pm^2} + \frac{\sqrt{m_2} \pm \sqrt{m_1}}{2\sqrt{2}}\frac{\ddot{d}}{\Omega_\pm^2}\right)^2 \ .
\ee

\section*{References}
\bibliography{refs}

\end{document}